\documentclass[12pt,a4paper]{article}
\usepackage{graphicx}
\usepackage{amssymb}
\usepackage{cite}
\usepackage{epsfig}
\global\arraycolsep=2pt 
\input{epsf}
\usepackage{color}
\def\be{\begin{equation}}
\def\ee{\end{equation}}
\def\pat{\partial}
\def\bea{\begin{eqnarray}}
\def\eea{\end{eqnarray}}
\def\beb{\begin{eqnarray*}}
\def\eeb{\end{eqnarray*}}

\newtheorem{thm}{Theorem}
\newtheorem{defn}[thm]{Definition}

\makeatletter
\def\fmslash{\@ifnextchar[{\fmsl@sh}{\fmsl@sh[0mu]}}
\def\fmsl@sh[#1]#2{%
  \mathchoice
    {\@fmsl@sh\displaystyle{#1}{#2}}%
    {\@fmsl@sh\textstyle{#1}{#2}}%
    {\@fmsl@sh\scriptstyle{#1}{#2}}%
    {\@fmsl@sh\scriptscriptstyle{#1}{#2}}}
\def\@fmsl@sh#1#2#3{\m@th\ooalign{$\hfil#1\mkern#2/\hfil$\crcr$#1#3$}}
\definecolor{pp}{rgb}{0.6, 0.1, 1}
\makeatother

\begin{document}


\thispagestyle{empty}
\begin{titlepage}

\begin{flushright}
LMU-TPW 2003-03 
\end{flushright}

\boldmath
\begin{center}
  {\large {\bf Introduction to a Non-Commutative Version of the Standard Model 
  }}
\end{center}
\unboldmath
\vspace{0.8cm}
\begin{center}
{Michael Wohlgenannt}
\end{center}
\vskip 1em
\begin{center}
Sektion Physik, Universit\"at M\"unchen,\\
Theresienstra{\ss}e 37, 80333 M\"unchen, Germany

\vspace{1.2cm}
The XIV International Hutsulian Workshop, Oct 28 - Nov 2, 2002, Chernivtsi,
Ukraine

\vspace{.1cm}
\end{center}

\vspace{4cm}
\begin{abstract}
\noindent
This article provides a basic introduction to some concepts of non-commutative
geometry. The importance of quantum groups and quantum spaces is stressed.
Canonical non-commutativity is understood as an approximation to the quantum
group case. Non-commutative gauge theory and the non-commutative Standard Model
are formulated on a space-time satisfying canonical non-commutativity relations.
We use $*$-formalism and Seiberg-Witten maps.
\end{abstract}


\end{titlepage}


\noindent
In these lectures I want to give a basic introduction to the Non-Commutative
Standard Model advocated in \cite{Calmet:2001na}. The underlying mathematical ideas
shall be introduced with some care. Although the ideas are in a way not
necessary to understand the physics discussed in the late Chapters, they are
nevertheless vital in gaining a better understanding and a sound picture. Why
we discuss some important aspects of non-commutative geometry first. We will
especially discuss the case of quantum groups and quantum spaces. Quantum spaces
carry a (co-)representation of a quantum group. The quantum group describes the
symmetry of that space. The notion of symmetry is a very important one in
physics, and it can be generalised to some classes of non-commutative spaces.
However, we will only discuss models on spaces obeying so-called canonical 
non-commutativity relation, which
does not allow for a symmetry group. Classical Lorentz symmetry is broken, and 
there is no deformation of the symmetry present. In some sense, one can think of
the canonical case as an approximation to the quantum group case. In both cases,
the non-commutativity is characterised by a parameter $q$ or $\theta$, respectively. Of
course, a very important feature is that in the limit of vanishing
non-commutativity - $q\to 1$, $\theta\to 0$, respectively - we end up with the usual
commutative theory. In Chapter 2, we will discuss a special approach to non-commutative geometry,
namely $*$-products. One of the advantages of this approach is that the 
commutative limit is very transparent. In the third Chapter, we will discuss
gauge theory on canonically non-commutative space-time \cite{Madore:2000en}. The last
two Chapters are devoted to the Standard Model and its generalisation to
non-commutative space-time using techniques developed in Chapter 3.

\section{Non-Commutative Geometry}

Let me first try to give you some handwaving idea what picture we have in mind when we talk
of non-commutative (nc) geometry. Some examples will show, where
non-commutativity has already shown up in physics, and how these ideas might
be useful. After all these motivations, I want to formulate some aspects of nc
geometry mathematically. We will mainly be concerned with quantum groups and
quantum spaces.

\subsection{\label{sec1.1}What is Non-Commutative Geometry?}

As the word "non-commutative" says, the commutator of some elementary quantities
is doomed not to vanish. Nc geometry is based on non-commutative coordinates
\be
{}[\hat x^i, \hat x^j ] \ne 0,
\ee
i.e., coordinates are non-commutative operators and we have to think in
quantum mechanical terms. $\hat x^i$'s cannot all be diagonalised 
simultaneously.
Space-time is the collection of the eigenvalues (spectrum) of the operators 
$\hat x^i$. If the spectrum is discrete, space-time will be discrete.
Commutative coordinates induce a continuous spectrum. Therefore, also
space-time will be continuous.

\noindent
The theory of nc geometry is based on the simple idea of replacing ordinary
coordinates with non-commuting operators. We will see how this idea can be
formulated mathematically.

\subsection{\label{sec1.2}Physical Motivation for Non-Commuting Coordinates}  

But before we do so, let us consider some examples.

\subsubsection{Divergencies in QFT}

\begin{figure}[htp]
\begin{center}
\begin{picture}(100,40)
\put(50,20){\circle{40} }
\put(0,0){\line(1,0){100}}
\end{picture}
\end{center}
\caption{Loop contribution}
\label{fig1}
\end{figure}
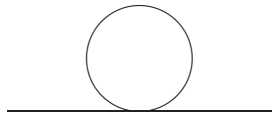

In quantum field theories, loop contributions to the transition amplitudes 
diverge.  
Consider a real scalar particle $\phi$ which is described by the
action
\be\label{phi^4}
\mathcal{L} = {1\over 2}\partial_\mu\phi\partial^\mu\phi - {1\over 2}m^2\phi^2 
+ {\lambda\over 4!}\phi^4.
\ee
The contribution of the diagram shown in Fig. \ref{fig1} reads
\be
\int {d^4\!q\over (2\pi)^4} \, {1\over q^2 - m^2}.
\ee
The result is divergent. The renormalisation procedure may remove some of the 
infinities. The
theory is called renormalisable, if all divergencies can be removed with a
finite number of counter terms. The theory
defined by (\ref{phi^4}) is renormalisable in $4$ dimensions. However,
no renormalisable quantum field theory of gravity is known so far. 
Discretising space-time may introduce a momentum cutoff in a
canonical way and render the theory finite or at least renormalisable.
The hope is that this can be accomplished by making space-time non-commutative.

\subsubsection{Quantum Gravity}

All kind of models for and approaches to quantum gravity seem to lead to a 
fundamental length scale, i.e., to a lower bound to any position measurement 
\cite{Garay:1995en}. This seems to be a model independent feature. The uncertainty in
space-time measurement can be explained by replacing coordinates by nc
operators.

\subsubsection{String Theory}

In open string theory with a background B-field, the endpoints of the 
strings are confined to submanifolds (D-branes) and become non-commutative
\cite{Seiberg:1999vs}. This is true even on an operator level,
\be
{}[X^i,X^j] = i\theta^{ij},
\ee
where $\theta^{ij}=-\theta^{ji}\in \mathbb{R}$, and $X^i$ are the coordinates 
of the $2$-dimensional world sheet embedded in the target space
(e.g., $\mathbb{R}^{10}$), i.e., operator valued bosonic fields. Therefore, we
also have for the propagator
\be
\langle \, [X^i,X^j] \, \rangle = i\theta^{ij}.
\ee

\subsubsection{Classical Non-Commuting Coordinates}

Consider a particle with charge $e$ moving in a homogeneous and constant 
magnetic field. The action is given by
\be
S=\int dt\left( {1\over 2}m\dot x_\mu \dot x^\mu - {e\over c} 
B_{\mu\nu}x^\mu\dot x^\nu
\right),
\ee
where $B_{\mu\nu}$ is an antisymmetric tensor defining the vector potential
$A_\mu$, $B_{\mu\nu}=-B_{\nu\mu}$ and $A_\nu = B_{\mu\nu}x^\mu$. The classical
commutation relations are
\be
\left\{ \pi_\mu, x^\nu \right\} = \delta^\nu_\mu,
\ee
where $\{\, , \,\}$ is the classical Poisson structure.
Writing it out explicitly, we get
\be\label{comm}
\left\{\dot x_\mu,x^\nu \right\} + {eB_{\mu\sigma}\over c\, m} 
\left\{ x^\sigma, x^\nu \right\} = {1\over m} \delta^\mu_\nu,
\ee
where $\pi_\mu={\partial \mathcal L \over \partial \dot{x}^\mu} = 
m\dot x_\mu + {e\over c} B_{\mu\nu} x^\nu$.
Let us assume strong magnetic field $B$ and small mass $m$ - i.e., we restrict 
the particle to the
lowest Landau level \cite{Landau:1930vs}. In this approximation,
eqn. (\ref{comm}) simplifies, and we get \cite{Jackiw:2001dj}
\be
\left\{ x^\sigma, x^\nu \right\} = {c \, (B^{-1})^{\sigma\nu}\over e}.
\ee
The coordinates perpendicular to the magnetic field do not commute, 
on a classical level.

\subsection{\label{sec1.3}Systematic Approach}

\begin{figure}[htp]
\begin{center}
\begin{picture}(350,70)
\put(10,30){\small compact }
\put(10,10){\small manifold $\mathcal{M}$}

\put(120,20){\vector(-1,0){30}}
\put(120,20){\vector(1,0){30}}

\put(165,20){\small $\mathcal{C}(\mathcal{M})$}
\put(105,35){\tiny Urysohn's}
\put(105,0){\tiny Lemma}

\put(240,20){\vector(-1,0){30}}
\put(240,20){\vector(1,0){30}}

\put(230,35){\tiny Gel'fand}
\put(230,0){\tiny Naimark}
\put(305,30){\small abelian, semisimple}
\put(305,10){\small $C^*$ algebra}

\end{picture}
\end{center}

\caption{Classical algebraic geometry}

\label{fig2}
\end{figure}
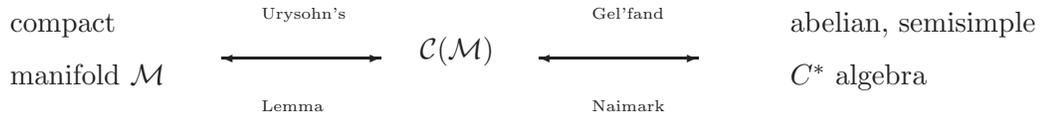

Let us examine the classical situation depicted in Fig. \ref{fig2}. 
We start with a smooth and compact manifold $\mathcal M$. The topology of
$\mathcal M$ is uniquely determined by the algebra of continuous complex (real)
valued functions on $\mathcal M$, $C(\mathcal M)$ with the usual involution 
(Urysohn's Lemma \cite{rudin}). The Gel'fand-Naimark theorem
\cite{gelfand} relates the function algebra to an abelian $C^*$-algebra. 
The algebra of continuous functions over a compact manifold $\mathcal M$ is
isomorphic to an abelian unital $C^*$-algebra. The algebra of continuous
functions vanishing at infinity over a locally compact Hausdorff space 
$C^0(\mathcal M)$ is isomorphic to an abelian $C^*$-algebra (not necessarily
unital).

\begin{figure}[htp]
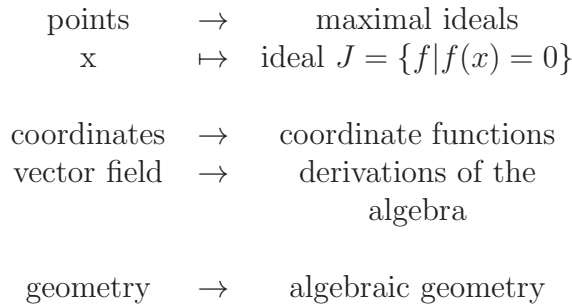


\begin{center}
\begin{tabular}{ccc}
  points & $\rightarrow$ & maximal ideals\\
  x      & $\mapsto$     & ideal $J=\{f|f(x)=0\}$\\
  \\
  coordinates & $\rightarrow$ & coordinate functions\\
  vector field& $\rightarrow$ & derivations of the\\
  &&algebra\\
  \\
  geometry & $\rightarrow$ & algebraic geometry
 \end{tabular}
\end{center}

\caption{Algebraic geometry}
\label{fig3}

\end{figure}

\noindent
Coordinates on the manifold are replaced by coordinate functions in 
$C(\mathcal M)$,
vector fields by derivations of the algebra. Points are replaced by maximal
ideals, cf.  Fig. \ref{fig3}.

\noindent
The trick in nc geometry is to replace the abelian $C^*$ algebra by a
non-abelian one and to reformulate as much of the concepts of algebraic geometry
as possible in terms of non-abelian $C^*$ algebras \cite{Madore:2000aq,connes1}.

\noindent
In the following, this nc algebra $\mathcal{A}$ will be given by 
the algebra  of formal power
series generated by the nc coordinate functions $\hat x^i$, divided by an ideal
$\mathcal I$ of relations generated by the commutator of the coordinate functions,
\be  
\mathcal{A}={\mathbb{C} \langle\langle \hat x^1,...\hat x^n \rangle\rangle \over \mathcal I},
\ee
where $[\hat x^i,\hat x^j]\ne 0$. Most commonly, the commutation relations are 
chosen to be either constant or linear or quadratic in the generators.
In the canonical case the relations are constant,
\be\label{canonical}
[\hat x^i,\hat x^j] = i \theta^{ij},
\ee
where $\theta^{ij}\in\mathbb{C}$ is an antisymmetric matrix, 
$\theta^{ij}=-\theta^{ji}$. The linear or Lie algebra case
\be\label{linear}
[\hat x^i,\hat x^j] = i\lambda^{ij}_k\hat x^k,
\ee
where $\lambda^{ij}_k\in\mathbb{C}$ are the structure constants, basically has
been discussed in two different approaches, fuzzy spheres \cite{Madore:1992bw} and
$\kappa$-deformation \cite{Lukierski:1991pn, Majid:1994cy,Dimitrijevic:2003aa}. Last 
but not least, we have quadratic commutation relations
\be\label{quadratic}
[\hat x^i,\hat x^j] = ({1\over q}\hat R^{ij}_{kl}-\delta^i_l
\delta^j_k) \hat x^k \hat x^l,
\ee
where $\hat R^{ij}_{kl}\in\mathbb{C}$ is the so-called $\hat R$-matrix which will be
discussed in some detail later, corresponding to quantum groups. For a
reference, see e.g., \cite{reshetikhin, Lorek:1997eh}.

Let us discuss one specific approach to nc geometry in some more detail, namely
quantum groups. The main part of this talk will deal with the Standard Model on
canonical space-time. We will consider canonical spaces as an approximation in
some sense, to
quantum spaces. The advantage of
quantum spaces is that the concept of symmetry can be generalised to quantum
groups. Whereas canonical space-time does not allow for a generalised 
Lorentz symmetry.

\subsubsection{q-Deformed Case}

Classically, symmetries are described by Lie algebras or
Lie groups. Physical spaces are representation
spaces of its symmetry algebra - or respectively co-representations of the
function algebra over its symmetry group.
Therefore, we will introduce nc spaces as co-representation spaces
of some quantum group. The interpretation goes as follows:
Space-time is a continuum in the
low energy domain, at high energies - Planck energy or below - 
space-time undergoes a phase transition and becomes a "fuzzy" nc space. 
Therefore the symmetries are not broken, but deformed to a
quantum group.

\noindent
What is a quantum group?
Let us start with the function algebra over a classical Lie algebra 
$\mathcal F(\mathcal G)$. $\mathcal F(\mathcal G)$ is a Hopf algebra which will be 
defined in a
minute. Then there is a well defined transition from the classical function
algebra to the respective quantum group, $\mathcal F(\mathcal G)
\to \mathcal F(\mathcal G)_q$,
introducing the non-commutativity parameter $q\in\mathbb{C}$.
In the classical limit, $q\to 1$, we have to regain the classical situation.
This is the basic property of a deformation. 

As we mentioned before, the enveloping algebra of a Lie algebra and the 
function algebra over a Lie group 
are in a natural way Hopf algebras. Most importantly, q-deformation does not 
lead out of the category of Hopf algebras. Therefore we will now define the
concept of Hopf algebras and quantum groups.

\vspace{1cm}
\noindent
{\bf Hopf algebra.} A Hopfalgebra $A$ (see, e.g., \cite{Klimyk:1997eb}) 
consists of an algebra and a co-algebra
structure which are compatible with each other. Additionally, there is a
map called antipode, which corresponds to the inverse of a group.
$A$ is an algebra, i.e., there is a
multiplication $m$ and a unit element $\eta$,
\beb
 m  & : &  A \otimes A  \to   A, \\ 
    &&     a \otimes b \mapsto a b,\\
\eta & : & \mathbb{C} \to A, \\
    && c \mapsto c\, \mathbf{1}_A,
\eeb
such that the multiplication satisfies the associativity axiom (Fig. \ref{fig4})
\begin{figure}[htp]
\begin{center}

\begin{picture}(400,100)

\put(10,90){$A\otimes A \otimes A$}
\put(85,95){ \vector(1,0){60} }
\put(99,100){\small $m\otimes id$}
\put(170,90){$A\otimes A$}
\put(37,85){ \vector(0,-1){45} }
\put(53,60){\small $id \otimes m$}
\put(25,20){$A\otimes A$}
\put(85,25){ \vector(1,0){60} }
\put(180,20){$A$}
\put(182,85){ \vector(0,-1){45} }
\put(196,60){\small $m$}
\put(110,30){\small $m$}

\put(250,60){$\widehat{=} \qquad\,\,(a b) c = a (b c)$}

\end{picture}
\end{center}

\caption{Associativity}
\label{fig4}

\end{figure}
and $\eta$ satisfies the axiom depicted in Fig. \ref{fig5}.

\begin{figure}[htp]
\begin{center}
\begin{picture}(400,100)
\put(10,50){$\mathbb{C}\otimes A$}
\put(70,55){ \vector(1,0){40} }
\put(120,50){$A\otimes A$}
\put(80,65){\small $\eta\otimes id$}
\put(215,55){ \vector(-1,0){40} }
\put(225,50){$A\otimes\mathbb{C}$}
\put(132,40){ \vector(0,-1){25} }
\put(129,0){$A$}
\put(185,65){\small $id\otimes \eta$}
\put(140,28){\small $m$}
\put(25,37){ \vector(3,-1){85} }
\put(240,37){\vector(-3,-1){85}}
\put(75,25){ $\cong$ }
\put(175,25){ $\cong$ }

\put(250,10){$\widehat{=} \qquad \,\, \mathbf{1}_A \cdot a = a \mathbf{1}_A = a$}

\end{picture}

\end{center}

\caption{Unity axiom}
\label{fig5}
\end{figure}

\noindent
Reversing all the arrows in Figs. \ref{fig4} and \ref{fig5} and replacing $m$
by the co-product $\Delta$ and $\eta$ by the co-unit $\epsilon$ gives us the
axioms for the structure maps of a co-algebra. The co-product and the co-unit,
\beb
\Delta: A & \to & A\otimes A,\\
\epsilon: A & \to & \mathbb{C}
\eeb
are the dual to $m$ and $\eta$, respectively.
Compatibility between algebra and co-algebra structure means that the co-product
$\Delta$ and the co-unit $\epsilon$ are algebra homomorphisms,
\bea
\Delta(a b) & = & \Delta(a)\Delta(b), \\
\epsilon(a b) & = & \epsilon(a)\epsilon(b),
\eea
where $a,b\in A$.
The antipode $S:A\to A$ satisfies the axiom shown in Fig. \ref{fig6}.
It is an anti-algebra homomorphism.

\begin{figure}[htp]
\begin{center}

\begin{picture}(400,100)

\put(10,90){$A\otimes A$}
\put(100,95){ \vector(-1,0){50} }
\put(75,100){\small $\Delta$}
\put(115,90){$A$}
\put(22,85){ \vector(0,-1){45} }
\put(35,60){\small $id \otimes S$}
\put(10,20){$A\otimes A$}
\put(50,25){ \vector(1,0){50} }
\put(115,20){$A$}
\put(119,85){ \vector(0,-1){45} }
\put(132,60){\small $\eta\circ\epsilon$}
\put(75,30){\small $m$}
\put(135,95){\vector(1,0){50}}
\put(155,100){\small $\Delta$}
\put(205,90){$A\otimes A$}
\put(205,20){$A\otimes A$}
\put(185,25){\vector(-1,0){50}}
\put(220,85){\vector(0,-1){45}}
\put(155,30){\small $m$}
\put(233,60){\small $id \otimes S$}

\put(275,60){$\widehat{=} $}
\put(290,60){
       $\left\{
	\begin{array}{l} 
	m\circ (S\otimes id)\circ \Delta  \\
	= \eta\circ\epsilon\\
	= m\circ (id\otimes S)\circ\Delta
	\end{array}
       \right.$
       }

\end{picture}
\end{center}

\caption{Antipode axiom}
\label{fig6}

\end{figure}

\noindent
If $A$ is the algebra of functions over some  matrix group, the antipode $S$ 
is the inverse, 
\be
S(t^i_j) = (t^{-1})^i_j,
\ee
where $t^i_j$ are the coordinate functions and generate the algebra of functions.

\vspace{.5cm}
\noindent
Let me quote the structure maps for the function algebra and its dual.
Let $\mathcal G$ be an arbitrary, (for simplicity) finite group and 
$\mathcal F(\mathcal
G)$ the Hopf algebra of all complex-valued functions on $\mathcal G$.
Then the algebra structure of $\mathcal F(\mathcal G)$ is given by
\bea
m & : & \mathcal F(\mathcal G)\otimes\mathcal F(\mathcal G) \to \mathcal 
	F(\mathcal G),\\
&& m(f_1\otimes f_2)(g) = f_1(g) f_2(g),\nonumber\\
\eta & : & \mathbb{C} \to \mathcal F(\mathcal G),\\
&& \eta(k) = k\, \mathbf{1}_{\mathcal F(\mathcal G)}.\nonumber
\eea
And we have the following co-algebra structure 
\bea
\Delta & : & \mathcal F(\mathcal G) \to \mathcal F(\mathcal G) \otimes 
	\mathcal F(\mathcal G),\\
&& \Delta(f)(g_1\otimes g_2) = f(g_1g_2),\nonumber\\
\epsilon & : & \mathcal F(\mathcal G) \to \mathbb{C},\\
&& \epsilon(f)  = f(e),\nonumber
\eea
where $e$ is the unit element of $\mathcal G$.
Eventually, the antipode is given by 
\be
(S(f))(g) = f(g^{-1}).
\ee 
$\mathcal F(\mathcal G)$ is a commutative Hopf algebra.

\vspace{.5cm}
\noindent
Let us consider its dual. Let {\small${\mathbf g}$} be a Lie algebra. 
The universal enveloping
algebra is defined as
\be
\mathcal U({\mathbf g}) = {T({\mathbf g})\over x\otimes y - y\otimes x - [x,y]},
\ee
where $T(g)$ is the universal tensor algebra.
Its algebra structure is given by the commutator and its unit element by the
unit in $T(\mathbf g)$. The other structure maps are consistently defined as
\bea
\Delta(x) & = & x\otimes {\mathbf 1} + {\mathbf 1}\otimes x,\\
\epsilon(x) & = & 0,\\
S(x) & = & -x. 
\eea
$\mathcal U(g)$ is a co-commutative Hopf algebra, i.e., the co-product is
symmetric.

\vspace{1cm}
\noindent
{\bf Quantum Group.} A quantum group is a Hopf algebra with one additional
structure. Let us concentrate on the function algebra $\mathcal F(\mathcal G)$
over some Lie group $\mathcal G$
rather than on its dual, the universal enveloping algebra of the Lie
algebra. The
additional structure is the $R$-form, 
$$
R: A\otimes A\to \mathbb{C}.
$$ 
$\mathcal F(\mathcal G)$ is a commutative algebra, the $R$-form
describes the almost commutativity of the deformed product. 
Let us denote the quantum group $\mathcal F(\mathcal
G)_q$, since $R$ depends on the non-commutativity parameter q. Let $t^i_j$ 
be the coordinate functions generating $\mathcal F(\mathcal G)_q$. The 
generators satisfy the so-called
$RTT$-relations expressing the almost commutativity
\be\label{rtt}
R^{ij}_{kl}\, t^k_m\, t^l_n = t^j_l\, t^i_k \, R^{kl}_{mn},
\ee
where $R(t^i_k\otimes t^j_l) = R^{ij}_{kl}$. In the commutative limit, $R\to
\mathbf{1}$, we have commutative generators,
\be
t^k_m\, t^l_n = t^l_n \, t^k_m.
\ee
$R$ is a solution of the Quantum-Yang-Baxter-Equation (QYBE)
\be\label{QYBE}
R_{12}R_{13}R_{23}=R_{23}R_{13}R_{12},
\ee
where $(R_{13})^{ijk}_{lmn}=\delta^j_m R^{ik}_{ln}$, $R_{12}$ and $R_{23}$ are
defined accordingly.

\vspace{.5cm}
\noindent
{\bf Quantum spaces, $\mathcal M_q$}. A quantum space for a quantum group
$\mathcal F(\mathcal G)_q$ has two basic properties. $\mathcal M_q$ is a 
$\mathcal F(\mathcal G)_q-$co-module algebra and in the commutative limit, 
$q\to 1$, $\mathcal M$ is the
proper $\mathcal F(\mathcal G)-$co-module space. 
\be
\mathcal M_q \equiv \mathbb{C} \langle\langle \hat x^1,\dots,\hat x^n
\rangle\rangle / \mathcal I,
\ee
where $\mathcal I$ is the ideal generated by the commutation relations of the
generators $\hat x^i$. But how can the commutation relations be chosen
consistently? The product in $\mathcal M_q$ has to be compatible with the 
co-action of $\mathcal F(\mathcal G)_q$.\\
First let us introduce $\widehat R\equiv R \circ \tau$. In the classical limit, $\widehat
R$ is just the permutation $\tau$, $\tau(a\otimes b)=b\otimes a$. 
$\widehat R$ can be decomposed into projectors,
\be
\widehat R = \lambda_1 \widehat P_S + \lambda_2 \widehat P_A,
\ee
where $\widehat P_A$ is the q-deformed generalisation of the antisymmetriser, $\widehat
P_S$ of the symmetriser, respectively. The relations 
\be\label{pa}
{\widehat P_A\,}^{mn}_{ij} \hat x^i\hat x^j = 0 
\ee
on $\mathcal M_q$ satisfy both requirements. In the commutative limit,
(\ref{pa}) means
that the commutator of two coordinates vanishes. It is also covariant under the
co-action $\rho$ of the quantum group
\bea
\rho:\mathcal M_q & \to & \mathcal F(\mathcal G)_q \otimes \mathcal M_q,\\
\rho(\hat x^i) & = & t^i_j \otimes \hat x^j,
\eea
since
\be
{\widehat P_A\,}^{mn}_{ij} (t^i_k\otimes \hat x^k) (t^j_l\otimes \hat x^l) = 
t^m_i\, t^n_j \otimes  {\widehat P_A\,}^{ij}_{kl}\, \hat x^k \hat x^l = 0.
\ee
$\widehat P_A$ is a polynomial in $\widehat R$, and the $\widehat RTT$ relations (\ref{rtt}) 
can be applied ($R^{ij}_{kl}= \widehat R^{ji}_{kl}$).

\vspace{.5cm}
\noindent
{\bf Differentials, $\widehat \partial_A$}.
$\hat \partial_A$ satisfy the same commutation relations as the 
coordinates \cite{Ogievetsky:1992pn},
\be
{\widehat P_A\,}^{ij}_{kl}\, \hat \partial_i \hat\partial_j = 0.
\ee
This follows from the assumptions on the exterior derivative $d$. The exterior
derivative   $d=\xi^A \hat \partial_A$
shall have the same properties as in the classical case,
\bea
d^2 & = & 0,\\
d \hat x^A & = & \xi^A + \hat x^A d,
\eea
where the coordinate differentials $\xi^A$ are supposed to anticommute, i.e.,
\be
{\widehat P_S\,}^{AB}_{CD}\, \xi^C\xi^D = 0.
\ee
Consequently, the differentials satisfy a modified Leibniz rule
\be
\hat\partial_A(\hat f\hat g) = (\hat \partial_A \hat f)\hat g + {\mathcal
O_A}^B(\hat f)\,\partial_B\hat g, 
\ee
where the operator ${\mathcal O_A}^B$ is a homomorphism ${\mathcal
O_A}^B(\hat f\hat g)={\mathcal O_A}^C(\hat f)\, {\mathcal O_C}^B(\hat g)$.

\vspace{.5cm}
\noindent
I want to finish this Section on quantum groups and quantum spaces with a
popular two dimensional example, the Manin plane.


\noindent
{\bf Example: Manin Plane,} see e.g., \cite{schraml}.
 The Manin plane is generated by the two coordinates
$\hat x,\,\hat y$. The underlying symmetry is given by the quantum algebra 
$\mathcal U_q(sl_2)$.

\begin{itemize}

\item The coordinates satisfy 
\be
 \hat x\hat  y  = q  \hat y \hat x .
\ee

\item The differentials fulfill the same relation, except for some scaling factor,
\be
\hat \partial_x \hat \partial_y = {1\over q} \hat \partial_y \hat \partial_x.
\ee

\item The crossrelations compatible with the above structures are given by 
\bea
\hat \partial_x \hat x & = & 1 + q^2 \hat x \hat \partial_x + q\lambda \hat y
	\hat\pat_y,\\
\hat \partial_x \hat y & = & q \hat y \hat \partial_x,\\ 
\nonumber\\
\hat \partial_y \hat x & = & q \hat x \hat \partial_y,\\
\hat \partial_y \hat y & = & 1 + q^2 \hat y \hat \partial_y,
\eea
where $\lambda=q-{1\over q}$.

\item The symmetry algebra $\mathcal U_q(sl_2)$ is  generated by $T^+, T^-,T^3$, which
satisfy the following defining relations
\bea
\frac{1}{q}T^+T^--qT^-T^+ & = & T^3, \nonumber\\
q^2T^3T^+-\frac{1}{q^2}T^+T^3 & = & (q+\frac{1}{q})T^+, \\
q^2T^-T^3-\frac{1}{q^2}T^3T^- & = & (q+\frac{1}{q})T^-. \nonumber
\eea

\item The action of the generators on coordinates is given by
\bea\label{ncspinor}
T^3 \hat x & = & q^2\,\hat x T^3 - q\hat x,\nonumber\\
T^3 \hat y & = & {1\over q^2}\hat y T^3 + {1\over q}\hat y,\nonumber\\
\nonumber\\
T^+ \hat x & = & q \hat x T^+ + {1\over q} \hat y,\\
T^+ \hat y & = & {1\over q}\hat y T^+, \nonumber\\
\nonumber\\
T^- \hat x & = & q \hat x T^-,\nonumber\\
T^- \hat y & = & {1\over q}\hat y T^- + q \hat x.\nonumber
\eea


\end{itemize}
is defined via the ordinary integral introducing a weight function $\omega$,
$\omega(x,y)={1\over xy}$.
In the classical limit, the symmetry algebra fulfills the usual $sl_2$
relations,
\bea
{} [ T^+, T^- ] & = &    T^3,\nonumber \\
{} [ T^3, T^+ ] & = &  2 \, T^+, \\
{} [ T^-, T^3 ] & = &  2 \, T^-.\nonumber
\eea
Eqns. (\ref{ncspinor}) reduce to the usual action of the generators of angular
momentum on a spin-$1/2$ state.


\subsubsection{Canonical Case}

For most parts of the rest of my lectures, we will concentrate on the canonical 
case. Thus let
us summarize the most important properties. As discussed before, 
Minkowski space with canonical commutation relations does not allow for a
Lorentz symmetry. Only a translational symmetry is present.
Compared to the quantum group case or other more sophisticated examples,
calculations can be done more easily and more interesting models can be studied.
The commutator of two coordinates is a constant
\be
[\hat x^\mu, \hat x^\nu]=i\theta^{\mu\nu},
\ee
where $\theta^{\mu\nu} = -\theta^{\nu\mu} \in \mathbb{C}$.
The derivatives act on coordinates as in the classical case,
\be
[\hat \partial_\nu, \hat x^\mu] = \delta^\mu_\nu.
\ee
However, there are two consistent ways to define commutation relations of
derivatives. By observing that 
\be
\hat x^\mu - i \theta^{\mu\nu} \hat\partial_\nu
\ee
commutes with all coordinates $\hat x^\nu$ and all derivatives $\hat\pat_\nu$
one may assume that this expression equals some constant, $0$ say. Thus, we can 
define a derivative in terms of the coordinates (for invertible
$\theta$),
\be
\hat \partial_\mu = -i \theta^{-1}_{\mu\nu}\hat x^\nu.
\ee
The commutator of derivatives is given by
\be
[\hat \partial_\mu,\hat  \partial_\nu] = i(\theta^{-1})_{\mu\nu}.
\ee
The other possibility is to define
\be
\hat \partial_\mu \hat f= -i\theta^{-1}_{\mu\nu} [\hat x^\nu, \hat f]
\ee
which leads to 
\be
[\hat \partial_\mu,\hat  \partial_\nu] = 0.
\ee
The integral is given by the usual four dimensional integral over commutative
functions
\be
\int \hat f \hat g := \int d^4\! x\, f*g \, (x) = \int d^4\! x\, f(x)\, g(x).
\ee
In the next Section we will discuss what we mean by the map $f \to \hat f$,
mapping the function $f$ depending on commutative
coordinates $x^\mu$ onto the nc function $\hat f$, and by the product $*$.
 
All the necessary prerequisites for field theory with action integral are met.
But before we are going to turn to physics, to gauge theory, we will talk about
the $*$-product approach. In this approach the commutative limit is very 
transparent.


\section{\label{sec2}Star Products}

Let us consider the nc algebra of functions $\widehat{\mathcal A}$ on canonical Minkowski
space
\be
\widehat{\mathcal A} = {\mathbb{C} \langle\langle \hat x^1,...,\hat x^n
\rangle\rangle \over \mathcal{I}},
\ee
where $\mathcal{I}$ is the ideal generated by the commutation relations of the
coordinate functions, and the commutative algebra of functions
\be
\mathcal{A} = {\mathbb{C} \langle\langle x^1,...,x^n \rangle\rangle 
\over [x^i,x^j]} \equiv \mathbb{C} [[x^1,...,x^n]],
\ee
i.e., $[x^i,x^j]=0$. Our aim in this Section is to relate these algebras by an
isomorphism. Let us first consider the vector space structure of the algebras,
only. In order to construct a vector space isomorphism, we have to choose a 
basis (ordering) in $\widehat{\mathcal A}$ - satisfying the
Poincar\'{e}-Birkhoff-Witt property - e.g., the basis of
symmetrically ordered monomials, 
\be 
1,\quad \hat x^i, \quad {1\over 2}(\hat x^i\hat  x^j + \hat x^j \hat x^i), \quad
\dots.
\ee
Now we map the basis monomials in
$\mathcal{A}$ onto the according symmetrically ordered basis elements of 
$\widehat{\mathcal A}$ 
\bea\label{quantisation}
W : \mathcal{A} & \to & \widehat{\mathcal A},\nonumber\\
x^i & \mapsto & \hat x^i,\\
x^i x^j & \mapsto & {1\over 2}( \hat x^i \hat  x^j + \hat x^j \hat x^i)\equiv
	\,\, :\hat x^i \hat x^j:.\nonumber
\eea
The ordering is indicated by $:\, :$.
$W$ is an isomorphism of vector spaces. 
In order to extend {\small $W$} to an algebra isomorphism, we have to 
introduce a new non-commutative multiplication $*$ in $\mathcal{A}$. This 
$*$-product is defined by
\be\label{star}
W(f*g) := W(f) \cdot W(g) = \hat f\cdot \hat g,
\ee
where $f, g \in \mathcal{A}$, $\hat f, \hat g \in \hat \mathcal{A}$.
\be
(\mathcal{A},*) \cong (\widehat{\mathcal A},\cdot),
\ee
i.e., {\small $W$} is an algebra isomorphism.
The information on the non-commutativity of $\widehat{\mathcal A}$ is encoded
in the $*$-product.

\subsection{\label{sec2.1}Construction of a $*$-Product of Functions}

Let us choose symmetrically ordered monomials as basis in $\widehat{\mathcal A}$. The
commutation relations of the coordinates are
\be
[\hat x^i, \hat x^j] = i \theta^{ij}(\hat x),
\ee
where $\theta(\hat x)$ is an arbitrary expression in the coordinates $\hat x$, for
now. In just a moment we will discuss the special cases
(\ref{canonical} - \ref{quadratic}).
The Weyl quantisation procedure \cite{Weyl:1927vd,Wigner:1932eb} is given by 
the Fourier transformation,
\bea
\label{invfour}
\hat f=W(f) & = & {1\over (2\pi)^{n/2}} \int d^n\!k\, e^{ik_j \hat x^j} 
	\tilde{f}(k),\\
\label{four}
\tilde f(k) & = & {1\over (2\pi)^{n/2}} \int d^n\!x\, e^{-ik_j x^j} f(x),
\eea
where we have replaced the commutative coordinates by nc ones ($\hat x^i$) 
in the inverse Fourier transformation (\ref{invfour}). The exponential takes
care of the symmetrical ordering. Using eqn. (\ref{star}), we get
\be\label{star2}
W(f*g)={1\over(2\pi)^n} \int d^n\!k\,d^n\!p\,
e^{ik_i\hat x^i}e^{ip_j \hat x^j} \tilde f (k) \tilde g (p).
\ee
Because of the non-commutativity of the coordinates $\hat x^i$, 
we need the Campbell-Baker-Hausdorff (CBH) formula
\be
e^A e^B = e^{A+B +{1\over 2}[A,B] + {1\over 12}[[A,B],B] - 
{1\over 12} [[A,B],A] + \dots}.
\ee
Clearly, we need  to specify $\theta^{ij}(\hat x)$ in order to evaluate the CBH
formula.

\noindent
{\bf Canonical Case.} Due to the constant commutation relations, the CBH formula
will terminate, terms with more than one commutator will vanish,
\be
\exp(ik_i \hat x^i) \exp(i p_j \hat x^j) = \exp \left( i(k_i + p_i)\hat x^i - 
{i\over 2}k_i\theta^{ij}p_j \right).
\ee
Eqn. (\ref{star2}) now reads
\be
f*g\, (x) = \int d^n\!k d^n\!p \,e^{i(k_i+p_i)x^i-{i\over 2}k_i\theta^{ij}p_j}
\tilde f(k)\tilde g(p)
\ee
and we get for the $*$-product the Moyal-Weyl product \cite{Moyal:1949sk}
\be
f*g\, (x) = \exp(\, {i\over 2} {\partial\over \partial x^i}\, \theta^{ij}
\,{\partial\over\partial y^j})\, f(x) g(y)\Big|_{y\to x}\, .
\ee

\noindent
{\bf Lie Algebra Case.} The coordinates build a Lie algebra 
\be
[\hat x^i,\hat x^j]=i\lambda^{ij}_k \hat x^k,
\ee 
with structure constants $\lambda^{ij}_k$. In this case the CBH
sum will not terminate and we get
\be
\exp(ik_i \hat x^i) \exp(i p_j \hat x^j) = \exp \left( i(k_i + p_i)\hat x^i
	+{i\over 2} g_i(k,p)\hat x^i \right),
\ee
where all the terms containing more than one commutator are collected in 
$g_i(k,p)$. (\ref{star2}) becomes
\be
f*g\, (x) = \int d^n\!k d^n\!p \, e^{i(k_i+p_i)x^i+{1\over 2} g_i(k,p)x^i} \tilde
	f(k)\tilde g(p).
\ee
The symmetrically ordered $*$-product takes the form
\be
f*g\, (x) = e^{ {i\over 2} z^i\, g_i(-i{\partial\over \partial x^i},
	-i{\partial\over\partial y^j}) } 
		\, f(x) g(y)\Big|_{z\to x \atop y\to x}\, .
\ee
In general, it will not be possible to write down a closed expression for the
$*$-product, since the CBH formula can be summed up only for very few examples.

\noindent
{\bf q-Deformed Case.}
The CBH formula cannot be used explicitly, we have to use eqns. 
(\ref{quantisation}), instead.
Let us first write functions  as a power series in $x^i$, 
\be
f(x)=\sum_J c_J\,\, (x^1)^{j_1}\cdot \dots (x^n)^{j_n},
\ee
where $J=(j_1,\dots, j_n)$ is a multi-index.
In the same way, nc functions are given by power series in ordered monomials
\be
\hat f (\hat x) = \sum_J c_J\,\, :(\hat x^1)^{j_1}\cdot \dots (\hat x^n)^{j_n}:.
\ee
In a next step, we have to express the product of two ordered monomials in 
the nc coordinates again in terms of ordered monomials, i.e., we have to find
coefficients $a_K$ such that 
\be\label{21}
:(\hat x^1)^{i_1}\dots (\hat x^n)^{i_n}:\,\, 
:(\hat x^1)^{j_1}\dots (\hat x^n)^{j_n}:\,\, =  \sum_K a_K 
:(\hat x^1)^{k_1}\dots (\hat x^n)^{k_n}:.
\ee
Knowing the $a_K$, we know the $*$-product for monimials. It is simply given by
\be\label{22}
(\hat x^1)^{i_1}\dots (\hat x^n)^{i_n}\, * \, (\hat x^1)^{j_1}\dots
(\hat x^n)^{j_n} = \sum_K a_K (\hat x^1)^{k_1}\dots (\hat x^n)^{k_n}
\ee
using the same coefficients $a_K$ as in (\ref{21}). The whole procedure makes
use of the isomorphism $W$ defined in eqns. (\ref{quantisation}) and
(\ref{star}). In a
last step we generalise the above expression to functions $f$ and $g$,
and express the $*$-product in terms of ordinary derivatives on the functions 
$f$ and $g$, respectively. This merely amounts to replacing $q^{i_k}$ - where
$i_k$ refers to the power of the k$^\textrm{\small th}$ coordinate in (\ref{22}) - 
by the
differential operator $q^{x^k \pat_k}$, where no summation over $i_k$ is
implied.
For the better illustration, let us consider some examples.

\noindent
{\bf Examples}
\begin{itemize}
\item The Manin plane is always an eligible candidate.
$$
\hat x \, \hat y = q \, \hat y \, \hat x.
$$
We consider normal ordering, i.e., a normal ordered monomial has the form
\be 
:\hat y^3\, \hat x^2\, \hat y: \, = \hat x^2\, \hat y^4.
\ee
Following the above prescription, we end up at the following $*$-product
\cite{Madore:2000en}
\be
f*g\,(x,y) = q^{ - \tilde x {\partial\over\partial \tilde x} y {\partial \over 
\partial y} }  f(x,y)\, g(\tilde x,\tilde y)\Big|_{\tilde x \to x \atop
\tilde y \to y}\, .
\ee

\item As a further example we want to quote the $*$-products on 
$q$-deformed Euclidean space in $3$ dimensions and of the 
$q$-Minkowski space \cite{Wachter:2001gk,Bauer:2002nt}. 
These products are more involved than the one on the Manin plane, but still the
structures show many similarities.

In case of the q-deformed $3$ dimensional Euclidean space, the algebra of
functions is generated by the coordinates $\hat x^+,\hat x^3, \hat x^-$. Again,
we consider normal ordering,
$$
:(\hat x^3)^{i_3}\, (\hat x^+)^{i_+}\, (\hat x^-)^{i_-}: \, = 
(\hat x^+)^{i_+}\, (\hat x^3)^{i_3}\, (\hat x^-)^{i_-}.
$$
The $*$-product is quoted as 
\be
f*g = 
\sum_{i=0}^\infty\lambda^i  \frac{x_3^{2i}}{[[i]]_{q^4}!}\,\,q^{2(\hat{\sigma}_3 
\hat{\sigma}'_++\hat{\sigma}_-\hat{\sigma}'_3)}\left(D_{q^4}^-\right)^i\!\! 
f(x)\cdot \left(D_{q^4}^+\right)^i\!\! g(x')\Big|_{x'\to x}\, ,
\ee
where $D_q^A\! f(x)=\frac{f(x_A)-f(qx_A)}{x_A-qx_A}$ is the discrete Jackson 
derivative and
$q^{\hat{\sigma}_A}=q^{x_A\frac{\partial}{\partial x_A}}$, where no summation is
implied. In \cite{Wachter:2001gk}
the $*$-product is also calculated in the four dimensional case.
However, the q-deformed Minkowski space resembles a much more complicated
structure than $4$ dimensional Euclidean space. Therefore the $*$-product is
much more involved, too. Again, we will only quote the result and will not
bother about the details. The coordinates are $\hat x^0, \hat x^+, \hat{\tilde
x}^3, \hat x^-$, normal odering is defined by monomials of the form
$ \hat x^0 \hat x^+ \hat{\tilde x}^3 \hat x^-$.
The $*$-product reads
\bea
f*g & = & \sum_{i=0}^\infty \left( \frac{\lambda_-}{\lambda_+} \right)^i 
	\sum_{k+j=i}  \frac{R_{k,j}\,(\underline{x})}{[[k]]_{q^2}!\,\, 
	[[j]]_{q^2}!}\,\, q^{(2\hat\sigma_3+\hat\sigma_-+i)\hat\sigma'_+ 
 	+(2\hat\sigma'_3+\hat\sigma'_++i)\hat\sigma_-}\times\\ 
& \times & \left[ (D^-_{q^2})^i\! f\right] \!\! (x_0,x_+,\tilde{x}_3,
	q^{j-k}x_-)\cdot \left[ (D^+_{q^2})^i\! g\right]\!\! 
	(x'_0,q^{j-k}x'_+,\tilde{x}'_3,x'_-)\Big|_{x'\to x} \, ,
\eea
where $R_{k,j}\,(\underline{x})= R_{k,j}\,(x^0, x^+, \tilde{x}^3, \hat x^-)$ is 
some polynomial in the coordinates.

\end{itemize}


\subsection{\label{sec2.2}Mathematical Approach to $*$-Products}

\begin{defn}{(Poisson Bracket)}\\
Let $\mathcal{M}$ be a smooth manifold, a Poisson bracket is a bilinear map
$\{,\}:\mathcal{C}^\infty(\mathcal{M}) \times \mathcal{C}^\infty 
(\mathcal{M}) \to \mathcal{C}^\infty(\mathcal{M}) $ satisfying
\begin{itemize}
\item[] $f,g,h\in\mathcal{C}^\infty(\mathcal M )$
\item $\{f,g\} = -\{g,f\}$, antisymmetry
\item $\{\{f,g\},h\}+\{\{g,h\},f\}+\{\{h,f\},g\} = 0$, Jacobi identity
\item $\{f,gh\} = \{f,g\}h + g\{f,h\}$, Leibniz rule 
\end{itemize}
\end{defn}

\noindent
Locally, we can always write the Poisson bracket with the help of a
antisymmetric tensor
\be
\{f,g\} = \theta^{ij}\, \partial_i f \partial_j g,
\ee
where $\theta^{ij}=-\theta^{ji}$. Because of the Jacobi identity
$\theta^{ij}$ has to satisfy 
\be
\,\theta^{ij}\partial_j
\theta^{kl} + \theta^{kj}\partial_j \theta^{ik} = 0.
\ee

\begin{defn}{($*$-Product)}\\
Let $f,g\in\mathcal{C}^\infty(\mathcal{M})$ and 
$C_i:\mathcal{C}^\infty(\mathcal{M})\times \mathcal{C}^\infty(\mathcal{M}) \to
\mathcal{C}^\infty(\mathcal{M})$ be local bi-differential operators. Then we
define the $*$-product $*:\mathcal{C}^\infty(\mathcal{M}) \times
\mathcal{C}^\infty(\mathcal{M}) \to \mathcal{C}^\infty(\mathcal{M})[[h]]$, by
\be\label{starproduct}
f*g = \sum_{n=0}^\infty h^n C_n(f,g),
\ee
such that the following axioms are satisfied:
\begin{itemize}
\item $*$ is an associative product.
\item $C_0(f,g)=f\, g$, classical limit.
\item ${1\over h}[f{*\atop ,} g] = -i\{f,g\}$, in the limit $h\to 0$,
semiclassical limit.
\end{itemize}
\end{defn}

\noindent
The rhs. of definition (\ref{starproduct}) is an element of 
$\mathcal{C}^\infty(\mathcal M)[[h]]$,
the algebra of formal power series in the formal parameter $h$ with coefficients
in $\mathcal{C}^\infty(\mathcal M)$. Therefore we can 
generalise the given definition of the $*$-product to a $\mathbb{C}[[h]]$-linear 
product in $\mathcal{B}=\mathcal{C}^\infty(\mathcal M)[[h]]$ by
\bea
\big( \sum_n f_n h^n \big) * \big( \sum_m g_m h^m\big) & = &
	\sum_{k,l}f_k g_l h^{k+l} \nonumber\\
& + & \sum_{k,l\ge 0, m\ge 1}C_m(f_k,g_l) h^{k+l+m}.
\eea

\begin{thm}{(Theorem by M. Kontsevich \cite{Kontsevich:1997vb})}\\ 
$*$-products exist for any Poisson structure on $\mathbb{R}^n$. 
He also gives an explicit construction for the functions $C_n$ in 
(\ref{starproduct}).
\end{thm}

\noindent
Changing the ordering in the non-commutative algebra leads to {\it gauge
equivalent} $*$-products. The gauge transtransformation of the $*$-product
is given by 
\be
\mathcal{D}f*\mathcal D g = \mathcal D (f*'g),
\ee
where 
\be
\mathcal D f = f +\sum_{n\ge 1} (ih)^n\, \mathcal D_n(f).
\ee
$\mathcal D_n$ is an differential operator of order $n$.

\vspace{2cm}
\noindent
In the following we stick to the canonical case of commutation relations,
$$
[\hat x^i, \hat x^j] = i\theta^{ij}, \,\, i,j=1,...,4,
$$
with the Weyl-Moyal product of functions
$$
f*g\, (x) = \exp(\, {i\over 2} {\partial\over \partial x^i}\, \theta^{ij}
\,{\partial\over\partial y^j})\, f(x) g(y)\Big|_{y\to x}.
$$
Furtheron, we have classical derivatives
$$
[\partial_i,\partial_j]=0,
$$
and we use the ordinary integral for the integration
\beb
\int f*g & = & \int d^4 \! x \, (f*g)(x) = \int d^4 \! x f(x) g(x),\\
\int f*g*h & = & \int g*h*f = \int (g*h)\cdot f.
\eeb


\section{\label{sec3}Gauge Theory on NC Spaces}

We will now concentrate on physics. We want to discuss the Standard Model on a
canonically deformed space-time. Before we can do so, we have to think about
gauge theory on non-commutative spaces, in general. Let us first briefly recall
classical gauge theory. We will discuss in some detail the features that are
essential for the nc generalisation.

\subsection{\label{sec3.1}Gauge Theory on Classical Spaces}

Internal symmetries are described by Lie groups or Lie algebras, respectively. The 
elements $T^a$ 
\be
[T^a,T^b]=f^{ab}_c T^c
\ee
are generators of the Lie algebra, where $f^{ab}_c$ are its structure constants. 
Fields are given by
$n$-dimensional vectors carrying a irreducible representation of the gauge 
group. Elements of the symmetry algebra are represented by
$n\times n$ matrices.
The free action of the field $\psi$ is given by
\be
\mathcal S = \int d^4\! x \mathcal{L} = \int d^4\! x \, \partial_\mu \psi \, 
\partial^\mu \psi.
\ee
Requiring the gauge invariance of the action $\mathcal S$, one has to introduce
additional fields, gauge fields and to replace the usual derivatives by covariant
derivatives $\mathcal D_\mu$.

\noindent
Let us start with the field $\psi$ building an irreducible representation of 
the gauge group, i.e.,
\be
\delta\psi(x) = i\alpha(x)\psi(x),
\ee
where $\alpha$ is Lie algebra valued, 
$$
\alpha(x) = \alpha_a(x) T^a.
$$
Observe that the derivative of a field $\psi$ does not transform covariantly, i.e.,
\be\label{sit1}
\delta \partial_\mu \psi\ne i\alpha(x)\partial_\mu\psi(x).
\ee
Replacing the usual derivatives $\partial_\mu$ by covariant derivatives 
$\mathcal D_\mu$ and demanding 
that  $\mathcal D_\mu\psi$ transforms covariantly, 
one has to introduce a gauge potential $A_\mu(x)$,
\beb
\mathcal D_\mu & = & \partial _\mu - ig A_\mu(x),\\
A_\mu(x) & = & A_{\mu a}(x)T^a,\\
\delta A_\mu(x) & = & {1\over g}\partial_\mu \alpha (x) + [\alpha (x),A_\mu(x)].
\eeb
As it is well known, the interaction fields are a consequence of the gauge 
invariance 
of the action. Interactions are gauge interactions.
The modified action reads
\be
\mathcal S=\int d^4\! x\, \mathcal D_\mu \psi \mathcal D^\mu \psi,
\ee
including gauge Fields $A_\mu$. Forgetting about mass terms, we still need a 
kinetic term for the gauge 
fields in our action.
The only requirements are the gauge invariance of the kinetic term and 
renormalisablility of the theory. That fixes the 
kinetic term uniquely. This is a crucial point, and the situation will be
different in the case of the NCSM. The action is given by
\be
\mathcal S = \int d^4\! x\, \left( \mathcal D_\mu \psi \mathcal D^\mu \psi +
Tr\, F_{\mu\nu}F^{\mu\nu} \right),
\ee
where $F_{\mu\nu}=\partial_\mu A_\nu - \partial_\nu A_\mu - ig [ A_\mu,
A_\nu]$ is the field strength. 
Considering abelian gauge symmetry, commutators in 
$F_{\mu\nu}$ and in $\delta A_\mu$ will vanish. Let us make one more important
remark: there is  a sharp distinction between internal and external symmetry 
transformations. As we will see, that is not true in the case of nc gauge
theory.

\subsection{\label{sec3.2}Nc Gauge Theory}

Non-commutative gauge theory, as presented in \cite{Madore:2000en, Jurco:2001rq}, is based on essentially
three principles,
\begin{itemize}

\item Covariant coordinates,
\item Locality and classical limit,
\item Gauge equivalence conditions.

\end{itemize}
Let us first briefly recall our starting point. We have non-commutative
coordinates
\beb
[\hat x^\mu, \hat x^\nu] & = & i\theta^{\mu\nu},\\
\hat \mathcal A & = & {\mathbb{C} \langle\langle \hat x^1,...,\hat x^n
\rangle\rangle \over \mathcal I},
\eeb
the product of function $f,g\in \mathcal A$ is given by the Weyl-Moyal product.

\subsubsection{Covariant Coordinates}

Let $\psi$ be a non-commutative field, i.e., $\hat \psi \in \oplus_{i=1}^n
\widehat{\mathcal A}$,
\be
\widehat \delta \widehat \psi (\hat x) = i\widehat \alpha \widehat \psi(\hat x)
\ee
or
\be
\widehat\delta\psi(x) = i\alpha * \psi(x),
\ee
in the $*$ formalism, where $W(\alpha)=\widehat \alpha$.
Now, a similar situation arises as in eqn. (\ref{sit1}), only the derivatives are
replaced by coordinates. The product of a field and a coordinate does not
transform covariantly, since the $*$-product is not commutative,
\be\label{sit2}
\widehat\delta (x*\psi(x)) = i\, x * \alpha(x) * \psi(x)
\ne  i\, \alpha(x) * x * \psi(x).
\ee
The arguments are the same as before, and we 
introduce covariant coordinates
\be
X^\mu  \equiv  x^\mu + A^\mu,
\ee
such that
\be
\widehat \delta(X^\mu*\psi)=i\alpha*(X^\mu*\psi).
\ee
The coovariant coordinates and the gauge potential transform under a nc gauge
transformation in the following way
\bea
\widehat \delta X^\mu & = & i[ \alpha\stackrel{*}{,}X^\mu ],\\
\widehat \delta A^\mu & = & i[ \alpha \stackrel{*}{,} x^\mu ] + i[\alpha \stackrel{*}{,} 
A^\mu].
\eea
Other covariant objects can be constructed from covariant coordinates, such as a
generalisation of the field strength,
\be
F^{\mu\nu} = [ X^\mu \stackrel{*}{,} X^\nu ] - i\theta^{\mu\nu},\qquad
\widehat \delta F^{\mu\nu} = i [\alpha \stackrel{*}{,} F^{\mu\nu}].
\ee
For non degenerate $\theta$, we can define another gauge potential
$V_\mu$
\bea
\label{a.1}
\widehat \delta V_\mu & = & \partial_\mu \alpha + i[\alpha \stackrel{*}{,} V_\mu],\\
\label{a.2}
F_{\mu\nu} & = & \partial_\nu V_\mu - \partial_\nu V_\nu -i[V_\mu
\stackrel{*}{,}V_\nu],\\
\label{a.3}
\widehat \delta F_{\mu\nu} & = & i [\alpha\stackrel{*}{,} F_{\mu\nu}],
\eea
using
\bea
& A^\mu=\theta^{\mu\nu} V_\nu, \,\, F^{\mu\nu}=i \theta^{\mu\sigma}\theta^{\nu\tau}
F_{\sigma\tau}, &\\
&i\theta^{\mu\nu}\partial_\nu f = [x^\mu\stackrel{*}{,}f].&\nonumber
\eea
And we get for the covariant derivatives
\bea
\mathcal D_\mu \psi & = & (\partial_\mu - i V_\mu)*\psi,\\
\widehat \delta \mathcal (D_\mu * \psi) & = & i\alpha * \mathcal D_\mu
\psi.\nonumber
\eea
Even for abelian gauge groups, the $*-$commutators in eqns. (\ref{a.1}) (\ref{a.2}) 
do not vanish, and the
theory has similarities to a non-abelian gauge theory on a commutative
space-time.

Let us have a closer look at the gauge parameters and the gauge fields. In
classical theory, the gauge parameter and the gauge field are Lie algebra
valued, as we have mentioned before. Two subsequent nc gauge transformations
are again a gauge transformation,
\be\label{a.4}
\delta_\alpha \delta_\beta - \delta_\beta \delta_\alpha =
\delta_{-i[\alpha,\beta]},
\ee  
where $-i[\alpha,\beta]=\alpha^a\beta_b f^{ab}_c T^c$.
However, there is a remarkable difference to the non-commutative case.
Let $M^\alpha$ be some matrix basis of the enveloping algebra of the internal
symmetry algebra. We can expand the gauge parameters in terms of this basis,
$\alpha=\alpha_a M^a,\,\,\beta=\beta_b M^b$.
Two subsequent gauge transformations take again the form
\be
\widehat \delta_\alpha \widehat\delta_\beta - \widehat\delta_\beta \widehat \delta_\alpha = 
	\widehat\delta_{-i[\alpha,\beta]},
\ee
but the commutator of two gauge transformations involves the $*$-commutator of
the gauge parameters, and
\be\label{a.5}
[\alpha\stackrel{*}{,}\beta] = {1\over
2}\{\alpha_a\stackrel{*}{,}\beta_b\}[M^a,M^b] + {1\over
2}[\alpha_a\stackrel{*}{,}\beta_b]\{M^a,M^b\},
\ee
where $\{M^a,M^b\}=M^aM^b+M^bM^a$ is the anti-commutator.
The difference to (\ref{a.4}) is the anti-commutator $\{M^a,M^b\}$, respectively the
$*$-commutator of the gauge parameters, $[\alpha_a\stackrel{*}{,}\beta_b]$. This
term causes some problems. Let us assume that $M^\alpha$ are the Lie algebra
generators. Does the relation (\ref{a.5}) close? Or does (\ref{a.5}) rule out
Lie algebra valued gauge parameters? Clearly, the only crucial term is the
anti-commutator. The anti-commutator of two hermitian matrices is again
hermitian. But the anti-commutator of traceless matrices is in general not
traceless. Relation (\ref{a.5}) will be satisfied for the generators of the fundamental
representation of $U(n)$. Therefore it has been argued \cite{Chaichian:2001mu}
that $U(n)$ - and with some difficulty SO(n) and Sp(n) \cite{sosp} - is the only
gauge group that can be generalised to nc spaces. But in fact arbitrary gauge
groups can be tackled. But the gauge parameters $\alpha,\beta$ and the gauge
fields $A_\mu$ have to be
enveloping algebra valued \cite{Madore:2000en, Jurco:2000ja}, in general. Gauge fields
and parameters now depend on infinitely many parameters, since the enveloping
algebra is infinite dimensional. Luckily, the infinitely many degrees of freedom
can be reduced to a finite number, namely the classical parameters, by the
so-called Seiberg-Witten maps we will discuss in the next paragraph.

\subsubsection{\label{seclocality}Locality and Classical Limit}

The nc $*$-product can be written as an expansion in a formal parameter $h$,
$$
f * g = f\cdot g + \sum_{n=1}^\infty h^n C_n(f,g).
$$
In the commutative limit $h\to
0$, the $*$-product reduces to the pointwise product of functions.
One may ask, if there is a similiar commutative limit for the fields?
The solution to this question was given for abelian gauge groups by \cite{Seiberg:1999vs},
\bea\label{a.6}
\widehat A_\mu[A] & = & A_\mu + {1\over 2}\theta^{\sigma\tau} \left(
	A_\tau\partial_\sigma A_\mu + F_{\sigma\mu} A_\tau\right) 
	+\mathcal O(\theta^2),\\
\label{a.7}
\widehat\psi[\psi, A] & = & \psi + {1\over 2}\theta^{\mu\nu}A_\nu\partial_\mu \psi
	+\mathcal O(\theta^2),\\
\label{a.8}
\widehat\alpha & = & \alpha + {1\over 2}\theta^{\mu\nu}A_\nu\partial_\mu\alpha
	+\mathcal O(\theta^2).
\eea
In this case, $\theta$ is the non-commutativity parameter.
First of all, let me introduce an important {\bf convention} to which we will 
stick 
from now on. Quantities with "hat" ($\widehat\psi,\,\widehat A,\,\widehat \alpha\dots
\in(\mathcal A,*)$) refer to non-commutative fields and gauge parameters, respectively 
which can be expanded (cf. above) in terms of the ordinary commutative fields
and gauge parameters, resp. ($\psi,\, A,\alpha$). 

\vspace{.3cm}
\noindent
The Seiberg-Witten maps (\ref{a.6} - \ref{a.8}) reduce the infinitely many
parameters of the enveloping algebra to the classical gauge parameters.

\noindent
The origins of this map are in string theory.
It is there that gauge invariance depends on the regularisation scheme
applied \cite{Seiberg:1999vs}.
Pauli-Villars regularisation provides us with classical gauge invariance
\be
\delta a_i=\partial_i\lambda,
\ee
whence point-splitting regularisation comes up with nc gauge invariance
\be
\widehat \delta \widehat A_i = \partial_i \widehat \Lambda 
+i[\widehat\Lambda \stackrel{*}{,}\widehat A_i].
\ee
Seiberg and Witten argued that consequently there must be a local map 
from ordinary gauge theory to non-commutative gauge theory
\be
\widehat A[a], \,\, \widehat\Lambda[\lambda,a]
\ee
satisfying
\be\label{a.9}
\widehat A[a+\delta_\lambda a] = \widehat A[a]+\widehat \delta_\lambda
\widehat A[a].
\ee
The Seiberg-Witten maps are solutions of (\ref{a.9}). By locality we mean that
in each order in the non-commutativity parameter $\theta$ there is only a finite number 
of derivatives.

\subsubsection{Gauge Equivalence Conditions}

Let us remember that we consider arbitrary gauge groups. 
Nc gauge fields $\widehat A$ and gauge parameters $\widehat \Lambda$ are enveloping algebra
valued. Let us choose a symmetric basis in the algebra, 
$T^a,\, {1\over 2}(T^aT^b+T^bT^a),\,\dots$, such that
\bea
\widehat \Lambda(x) & = & \widehat \Lambda_a(x)T^a + \widehat \Lambda^1_{ab}(x)\,:T^aT^b: + 
	\dots\, , \\
\widehat A_\mu(x) & = & \widehat A_{\mu a}(x) T^a + \widehat A_{\mu ab}(x)\, :T^aT^b: 
	+ \dots\, .
\eea
Eqn. (\ref{a.9}) defines the SW maps for the gauge field and the gauge
parameter. However, it is more practical to find equations for the gauge 
parameter and the
gauge field alone \cite{Jurco:2001rq}.
First we will concentrate on the gauge parameters $\widehat\Lambda$. We already 
encountered the consistency condition
$$
 \widehat \delta_\alpha \widehat\delta_\beta - \widehat\delta_\beta \widehat
 \delta_\alpha = \widehat\delta_{-i[\alpha, \beta]}.
$$
More explicitly, it reads
\be\label{a.10}
i\widehat\delta_\alpha \widehat\beta[A] - i\widehat\delta_\beta
\widehat\alpha[A] + [\widehat\alpha[A]\stackrel{*}{,}\widehat\beta[A]] = 
(\widehat{[\alpha,\beta]})[A].
\ee
Keeping in mind the results from Section \ref{seclocality}, we
can expand $\widehat\alpha$ in terms of the non-commutativity $\theta$,
\be
\widehat\alpha[A] = \alpha + \alpha^1[A] +\alpha^2[A]+\dots,
\ee
where $\alpha^n$ is $\mathcal O(\theta^n)$.
The consistency relation (\ref{a.10}) can be solved order by order in $\theta$.
\bea
0^{\textrm{th}} \textrm{ order}: \,\alpha^0 & = & \alpha, \\
1^{\textrm{st}} \textrm{ order}: \,\alpha^1 & =  & {1\over
4}\theta^{\mu\nu}\{\partial_\mu \alpha, A_\nu\}, \\
	& = & {1\over 2}\theta^{\mu\nu}\partial_\mu\alpha_a A_{\mu b} 
		:T^aT^b:.
\eea
For fields $\widehat \psi$ the condition
\be\label{a.11}
\delta_{\alpha}\widehat\psi[A] = \widehat\delta_{\alpha}\widehat \psi[A] =
	i\widehat\alpha[A]*\widehat\psi[A]
\ee
has to be satisfied, keeping in mind that $\delta_\alpha$ denotes an ordinary gauge
transformation and $\widehat \delta_{\alpha}$ a nc one. That means that the
ordinary gauge transformation induces a nc gauge transformation.
We expand the fields in terms of the non-commutativity
\be
\widehat\psi =  \psi^0 + \psi^1[A] + \psi^2[A] + \dots
\ee
and solve eqn. (\ref{a.11}) order by order in $\theta$. In first order, we have
to find a solution to
\be
\delta_\alpha\psi^1[A]=i\alpha\psi^1+i\widehat\alpha\psi
-{1\over 2}\theta^{\mu\nu}\partial_\mu\alpha\partial_\nu\psi.
\ee
It is given by
\bea
0^{\textrm{th}} \textrm{ order}: \, \psi^0 & = & \psi,\\
1^{\textrm{st}} \textrm{ order}: \, \psi^1 & = & -{1 \over
	2}\theta^{\mu\nu} A_\mu\partial_\nu\psi + {i\over 4}\theta^{\mu\nu}
	A_\mu A_\nu\psi.
\eea
The gauge fields $\widehat A_\mu$ have to satisfy
\be\label{a.12}
\delta_{\alpha}\widehat A_\mu[A] = 
\partial_\mu\widehat\alpha[A]+i[\widehat\alpha[A]\stackrel{*}{,}\widehat A_\mu[A]].
\ee
Using the expansion
\be
\widehat A_\mu[A] = A^0_\mu + A_\mu^1[A] + A_\mu^2[A] + \dots 
\ee
and solving (\ref{a.12}) order by order, we end up with
\bea
0^{\textrm{th}}\textrm{ order} : \, A^0_\mu  & = & A_\mu,\\
1^{\textrm{st}}\textrm{ order} : \, A^1_\mu  & = & -{1\over
4}\theta^{\tau\nu} \{A_\tau, \partial_\nu A_\mu + F_{\nu\mu} \},
\eea
where $F_{\nu\mu}=\partial_\nu A_\mu - \partial_\mu A_\nu -
i[A_\nu,A_\mu]$. Similarly, we have for the field strength $\widehat F_{\mu\nu}$
\bea
\delta_\alpha \widehat F_{\mu\nu} & = & 
	i[\widehat \alpha,\widehat F_{\mu\nu}] \mbox{ and }\\
\widehat F_{\mu\nu} & = & F_{\mu\nu} + {1\over 2}\theta^{\sigma\tau}
	\{F_{\mu\sigma},F_{\nu\tau}\}
 -{1\over 4}\theta^{\sigma\tau}\{A_\sigma, (\partial_\tau+\mathcal
	D_\tau)F_{\mu\nu}\},
\eea
where $\mathcal D_\mu F_{\tau\nu} = \partial_\mu F_{\tau\nu} -
i[A_\mu,F_{\tau\nu}]$.

\subsubsection{Remarks}

Let us conclude this Section with some remarks and observations.

\begin{itemize}

\item SW maps provide solutions to the gauge equivalence relations.

\item Gauge equivalence relations are not the only possibile approach
to SW maps. 
Another approach is via nc Wilson lines, see e.g., \cite{Liu:2000mj}.

\item However, a certain ambiguity in the SW map remains. They are unique
modulo classical field redefinition and nc gauge transformation.
We used these ambiguities in order to choose $\widehat\Lambda$, 
$\widehat A_\mu$ hermitian. The freedom in SW map may also be essential for
renormalisation issues. There, parameters characterising the freedom in the
SW maps become running coupling constants \cite{Bichl:2001cq}.
Discussing tensor products of gauge groups, this freedom will also be of crucial
importance in Section \ref{secncsm}.

\item Gauge groups in non-commutative spaces contain space-time translations.
Since
\be
\partial f = -i\theta^{-1}_{ij}[x^j,f],
\ee
we can express the translation of the field $A_i$ as 
\be
\delta A_i = v^j\partial_j A_i = i[\epsilon\stackrel{*}{,}A_i],
\ee
\label{ay}
where $\epsilon = - v^j \theta^{-1}_{jk} x^k$. The gauge transformation of 
$A_i$  with gauge parameter $\epsilon$ gives
$$
\hat \delta_\epsilon A_i = i[\epsilon\stackrel{*}{,}A_i] - v^j\theta^{-1}_{ji}.
$$
This agrees with (\ref{ay}), ignoring the overall constant, which has no
physical effect \cite{Douglas:2001ba}.

\item NC gauge theory allows the construction of realistic particle models 
on a nc space-time with an arbitrary gauge group as internal symmetry group.
Nc gauge parameters and gauge fields are enveloping algebra valued, in general
(e.g., for $SU(n)$), but via SW maps the
infinite number of degrees of freedom is reduced to the
classical gauge parameters. Therefore these models will have the proper number 
of degrees of freedom.

\end{itemize}



\section{\label{sec4}Standard Model of Theoretical Particle Physics}

The Standard Model of Particle Physics \cite{ssm} is a 
very successfull and experimentally very well tested theory. It unifies strong,
weak and electromagnetic interactions. The aim of this section is to give a very
brief sketch. The Standard Model is a gauge theory with gauge group
$SU(3)_C\times SU(2)_L \times U(1)_Y$.

\subsection{Particle Content}

The particle content of the Standard Model is given in Table \ref{table1}.
\begin{table}[htp]
\begin{center}
  \begin{tabular}{ccccc}
  \hline
  & $SU(3)_C$ & $SU(2)_L$ & $U(1)_Y$  & $U(1)_Q$ 
   \\
   \hline
     $ e_R$
   & ${\bf 1}$
   & ${\bf 1}$ 
   & $-1$
   & $-1$  \\
   $ L_L=\left(\begin{array}{c} \nu_L \\ e_L \end{array} \right )$
   & ${\bf 1}$
   & ${\bf 2}$ 
   & $-1/2$
   & $\left(\begin{array}{c} 0 \\ -1 \end{array} \right )$  \\
   $u_R$
   & ${\bf 3}$
   & ${\bf 1}$ 
   & $2/3$
   & $2/3$  \\
   $d_R$
   & ${\bf 3}$
   & ${\bf 1}$ 
   & $-1/3$
   & $-1/3$  \\
     $Q_L=\left(\begin{array}{c}  u_L \\ d_L \end{array} \right )$
   & ${\bf 3}$
   & ${\bf 2}$ 
   & $1/6$
   & $\left(\begin{array}{c} 2/3 \\ -1/3 \end{array} \right )$  \\
$\Phi=\left(\begin{array}{c}  \phi^+ \\  \phi^0 \end{array} \right )$
   & ${\bf 1}$
   & ${\bf 2}$ 
   & $1/2$
   & $\left(\begin{array}{c} 1 \\ 0 \end{array} \right )$  \\
$B^i$
   & ${\bf 1}$
   & ${\bf 3}$ 
   & $0$
   & $(\pm 1, 0)$  \\
$A$
   & ${\bf 1}$
   & ${\bf 1}$ 
   & $0$
   & $0$  \\
   $G^a$
   & ${\bf 8}$
   & ${\bf 1}$ 
   & $0$
   & $0$  \\
   \hline
 \end{tabular}
\end{center}
\caption{Particle content of the Standard Model}
\label{table1}
\end{table}

\noindent
The action is given by
\begin{eqnarray}
&&
S=\int d^4x \sum_{i=1,2,3} \overline{\Psi}^{(i)}_L  i
  {\fmslash D} \Psi^{(i)}_L
  +\int d^4x \sum_{i=1,2,3} \overline{\Psi}^{(i)}_R i
  {\fmslash  D}  \Psi^{(i)}_R \nonumber \\
&& 
  -\int d^4x \frac{1}{4 g'}f_{\mu \nu}f^{\mu \nu}
  -\int d^4x \frac{1}{2 g}\mbox{{\bf tr}}F^L_{\mu \nu}F^{L \mu \nu} 
  -\int d^4x \frac{1}{2 g_S} \mbox{{\bf tr}}F^S_{\mu \nu}F^{S\mu \nu} 
  	\nonumber\\
&& \label{4.1}
  + \int d^4x \bigg(  (D_\mu   \Phi )^\dagger D^\mu \Phi            
  - \mu^2 {\Phi}^\dagger  \Phi - 
  \lambda \Phi^\dagger\Phi\Phi^\dagger \Phi   \bigg) \\ 
&& 
  + \int d^4x \bigg ( 
  -\sum_{i,j=1}^3 \bigg(
  W^{ij} \, ( \bar{   L}^{(i)}_L \Phi)
     e^{(j)}_R
  + {W^\dagger}^{ij}  \bar {  e}^{(i)}_R \Phi^\dagger    
  L^{(j)}_L) \bigg )  \nonumber \\ 
&&
  -\sum_{i,j=1}^3  \bigg(
  G_u^{ij} ( \bar{  Q}^{(i)}_L  {\bar\Phi})   
  u^{(j)}_R
  + {G_u^\dagger}^{ij} \bar {  u}^{(i)}_R   
   {\bar\Phi}^\dagger Q^{(j)}_L) \bigg ) \nonumber \\ 
&&
  -\sum_{i,j=1}^3  \bigg(
  G_d^{ij} ( \bar{   Q}^{(i)}_L  \Phi ) d^{(j)}_R
  + {G_d^\dagger}^{ij} \bar{ d }^{(i)}_R \Phi^\dagger Q^{(j)}_L) \bigg ) 
    \bigg) \nonumber,
\end{eqnarray}
where
\be
\Psi^{(i)}_L= \left ( \matrix{ L^{(i)}_L \cr
 Q^{(i)}_L
    } \right),\qquad
 \Psi^{(i)}_R = \left ( \matrix{ e^{(i)}_R \cr
 u^{(i)}_R \cr   d^{(i)}_R
    } \right), \qquad 
    {\Phi} =
 \left(\begin{array}{c}  \phi^+ \\  \phi^0
   \end{array} \right ).
\ee
$\Psi_L$ denotes the left handed fermions, the leptons $L$ and the quarks $Q$,
$\Psi_R$ denotes the right handed fermions. 
$(i)\in\{1,2,3\}$ is the generation index, and $\phi^+$ and
$\phi^0$ are the complex scalar fields
of the scalar Higgs doublet. $g' \mathcal A(x)$ is the gauge field of the 
hypercharge symmetry, $U(1)_Y$, $B_\mu (x)={g\over 2}B_{\mu a}(x)\sigma^a$
the field of the weak interaction, $SU(2)_L$ and 
$G_\mu(x) = {g_S\over 2}G_{\mu a}(x)\lambda^a$ of the strong
interaction, $SU(3)_C$. 
Some exemplary covariant derivatives have the form
\bea
\mathcal D_\mu \Phi & = & \left( \partial_\mu - i{g'\over 4}\mathcal A_\mu
	-i{g_S\over 2} B_{\mu a}\lambda^a \right) \Phi,\\
\mathcal D_\mu Q^{(i)}_L & = & \left(  \partial_\mu - i{g'\over 12}\mathcal
	A_\mu -i{g\over 2}B_{\mu a}\sigma^a - i{g_S\over 2} G_{\mu b}\lambda^b 
	\right) Q^{(i)}_L,\\
F^L_{\mu\nu} & = & g\left( \partial_\mu B_\nu - \partial_\nu B_\mu - 
	ig[B_\mu,B_\nu] \right).
\eea
Fermions and bosons are massless, initially. Spontaneous breaking of
the electroweak gauge symmetry $SU(2)_L\times U(1)_Y$ is responsible for the
masses of the gauge bosons, cf. Section \ref{sec4.2}. The Yukawa coupling 
terms in (\ref{4.1}) give masses to the fermions.
$W^{ij}$, $G^{ij}_u$ and $G^{ij}_d$ in (\ref{4.1}) are the Yukawa couplings.

\subsection{\label{sec4.2}Spontaneous Symmetry Breaking}

\begin{figure}[htp]
\centerline{\epsfig{figure=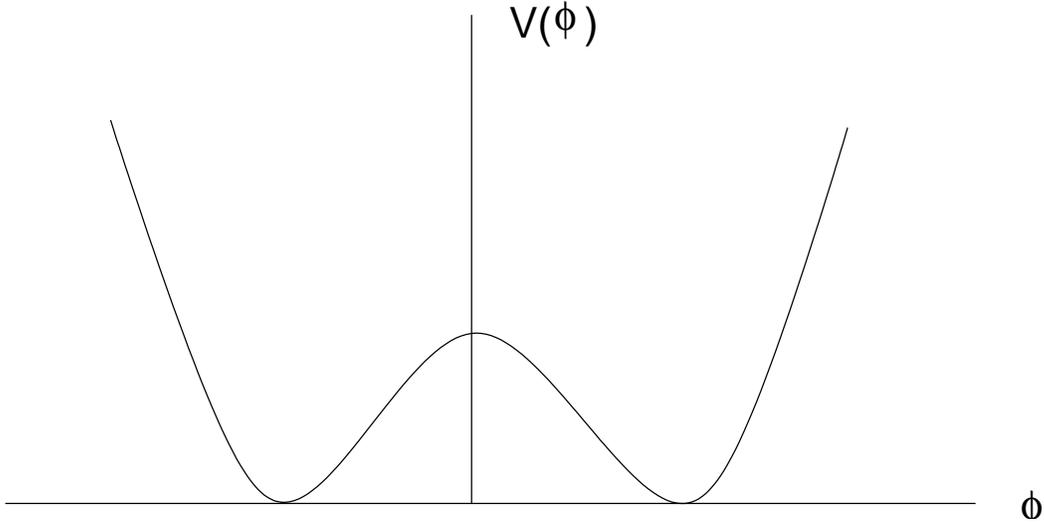}}
\caption{Mexican hat potential}
\label{fig7}
\end{figure}

Spontaneous symmetry breaking is characterised by two properties, the action is
invariant under a symmetry, but the solutions of the equations of motion are not.
A very popular example is the Mexican hat potential shown in Fig. \ref{fig7}. 
Let us consider the Higgs field. It has a $SU(2)\times U(1)$ gauge symmetry. The action is given
by
\be\label{d.2}
S = \int d^4 x\, \left( (D^\mu \phi)^\dagger D_\mu \phi - V(\phi) \right),
\ee
where the potential has the form
\be\label{d.1}
V(\phi) = \mu^2 \phi^\dagger \phi + \lambda (\phi^\dagger \phi)^2.
\ee
Let us write the complex Higgs field as
\be
\phi(x) = \left( \begin{array}{c} \phi_1(x) + i \phi_2(x) \\
	\phi_3(x) + i \phi_4(x) \end{array} \right),
\ee
where $\phi_1$, $\phi_2$, $\phi_3$ and $\phi_4$ are real fields.
If the parameter $\mu^2>0$ in (\ref{d.1}), (\ref{d.2}) describes four real particles with
mass $\mu$. In case of $\mu^2<0$, the potential $V$ resembles a Mexican hat potential
and spontaneous symmetry breaking occurs. The minima of the potential lie on circle around
$V(\phi=0)$, e.g., the point 
\bea\label{d.3}
& \phi_1 = \phi_2 = \phi_4 = 0,&\\
& \phi_3 = \sqrt{-\mu^2\over \lambda}& \nonumber
\eea
is an element of that circle. 
Each of these minima describes a possible vacuum state. Choosing one as the actual 
physical vacuum breaks the $SU(2)_L\times U(1)_Y$ gauge symmetry. Let us choose 
the point (\ref{d.3}) as the vacuum, and let us expand around that vacuum, and let us fix the gauge freedom to the unitarity gauge,
\be\label{d.4}
\phi(x) = {1\over \sqrt{2}} \left( \begin{array}{c} 0 \\ v + h(x) \end{array} \right),
\ee 
where $v=\sqrt{-\mu^2\over \lambda}$ is the vacuum expectation value. Expanding the action (\ref{d.2}) will only leave us with the kinetic term, since
\bea\label{d.5}
V(\phi) & = & V(\phi_0) + \Delta \phi {\pat V \over \pat \phi} \Big|_{\phi_0} + \mathcal 
	O(\Delta\phi^2)\\
& = & 0 + \mathcal O(\Delta \phi^2)\nonumber,
\eea
where the chosen vacuum is denoted by $\phi_0$.
The expansion of the kinetic term has the form
\bea
\label{d.6}
( D_\mu\phi )^\dagger ( D_\mu\phi ) & = & {1\over 2} ({vg\over 2})^2 B^+_\mu B^{- \mu} + 
	{1\over 8} v^2 
	\left( \begin{array}{cc}B^3_\mu & \mathcal A_\mu \end{array} \right)
	\left( \begin{array}{cc} g^2 & -gg´ \\ -gg´ & g´^2 \end{array} \right)
	\left( \begin{array}{c} B^{3 \mu} \\ \mathcal A^\mu \end{array} \right)\\
&& + \pat_\mu h\, \pat ^\mu h + \textrm{interaction terms}, \nonumber
\eea
where $B^\pm_\mu = {1\over \sqrt{2}} \left( B^1_\mu \pm i B^2_\mu \right)$. We can
identify $B^\pm$ with the $W^\pm$ bosons. As one can see from (\ref{d.6}), they are 
massive. Their mass is given by
\be\label{d.7}
M_{W^\pm} = {1\over 2} vg.
\ee
By diagonalising the matrix in (\ref{d.6}), we find the eigenvectors that correspond
to the other two particles that are still missing, namely the $Z$ boson and the photon.
Diagonalisation yields
\bea
\label{d.8}
Z_\mu & = & {gB^3_\mu - g´ \mathcal A_\mu \over (g^2 + g´^2)^{1/2} },\\
\label{d.9}
A_\mu & = & {g´ B^3_\mu + g\mathcal A_\mu \over (g^2 + g´^2)^{1/2} },
\eea 
with masses $M_Z={v\over 2} (g^2+g´^2)^{1/2}$ and $M_A=0$, respectively. The mass of the
electromagnetic photon is zero, that means that the $SU(2)_L\times U(1)_Y$ gauge 
symmetry is broken to $U(1)_{em}$.

\noindent
The Yukawa coupling terms in the action of the Standard Model supply the fermions with 
masses. The resulting terms are of the form, e.g.,
\be\label{d.10}
-{m_e\over 2} (\bar e_R e_L + \bar e_L e_R).
\ee



\section{\label{secncsm}Non-Commutative Standard Model}

This Section is based on \cite{Calmet:2001na}, a joint work with Xavier Calmet,
Branislav Jur\v{c}o, Peter Schupp and Julius Wess. 
We are going to deal with the symmetry group 
$SU(3)_C\times SU(2)_L\times U(1)$. Let me stress that we will therefore 
not introduce any new degrees of
freedom or any new parameters we would have to get rid off in the
end. The aim is not to reduce the number of parameters in the Standard Model - the Higgs
mass or parameters in the Higgs potential, the masses of the fermions or the Yukawa
coupling constants, etc. - but to formulate a realistic gauge theoretical model on
canonical space-time using the formalism introduced in Section \ref{sec3}. Especially 
interesting is whether the Higgs mechanism in the Standard Model can still be applied
to provide the boson masses.

Naively spoken,  the method we use is very simple. We just write down the Standard 
Model Lagrangian, replace $\cdot$ by $*$ and fields $\Psi$, $A$ by 
$\widehat\Psi[\Psi,A]$, $\widehat A[A]$. Of course, it is not as easy as that. I have
written down the non-commutative Lagrangian in (\ref{b.1}). In the rest of my
talk I am going to explain what all these terms mean. First of all, we have to
address some restrictions on non-commutative gauge theories discussed in
\cite{Chaichian:2001mu}. We have already discussed in Section \ref{sec3.2} the
problem of generalising other gauge groups than $U(n)$ to non-commutative
spaces. Further it is argued that a non-commutative field may be charged under
at most two gauge groups only. This problem will be solved in \ref{sec5.1} 
combining all the gauge fields of the Standard Model
into a single "master gauge field" und applying the SW map.
\begin{eqnarray}
S_{NCSM} & = & \int d^4x \sum_{i=1}^3 \overline{\widehat \Psi}^{(i)}_L \star i
  \widehat{\fmslash D} \widehat \Psi^{(i)}_L
  +\int d^4x \sum_{i=1}^3 \overline{\widehat \Psi}^{(i)}_R \star i
  \widehat{\fmslash  D} \widehat \Psi^{(i)}_R  \nonumber \\
&&-\int d^4x \frac{1}{2 g'} 
  \mbox{{\bf tr}}_{\bf 1} \widehat
  F_{\mu \nu} \star  \widehat F^{\mu \nu}
  -\int d^4x \frac{1}{2 g} \mbox{{\bf tr}}_{\bf 2} \widehat
  F_{\mu \nu} \star  \widehat F^{\mu \nu}
  -\int d^4x \frac{1}{2 g_S} \mbox{{\bf tr}}_{\bf 3} \widehat
  F_{\mu \nu} \star  \widehat F^{\mu \nu} \nonumber \\
&& + \int d^4x \bigg( \rho_0(\widehat D_\mu \widehat \Phi)^\dagger
  \star \rho_0(\widehat D^\mu \widehat \Phi) 
 -\mu^2 \rho_0(\widehat {\Phi})^\dagger \star  \rho_0(\widehat \Phi)  \nonumber
	\\
\label{b.1}
&& - \lambda
  \rho_0(\widehat \Phi)^\dagger \star  \rho_0(\widehat \Phi)\star
  \rho_0(\widehat \Phi)^\dagger \star  \rho_0(\widehat \Phi)   \bigg)  \\ 
&& + \int d^4x \bigg ( -\sum_{i,j=1}^3  \bigg(
  W^{ij} ( \bar{ \widehat L}^{(i)}_L \star \rho_L(\widehat \Phi))
  \star  \widehat e^{(j)}_R
  + {W^\dagger}^{ij} \bar {\widehat e}^{(i)}_R \star (\rho_L(\widehat \Phi)^\dagger \star \widehat
  L^{(j)}_L) \bigg )  \nonumber \\ 
&& -\sum_{i,j=1}^3  \bigg(
  G_u^{ij} ( \bar{\widehat Q}^{(i)}_L \star \rho_{\bar Q}(\widehat{\bar\Phi}))\star  
  \widehat u^{(j)}_R
  + {G_u^\dagger}^{ij} \bar {\widehat u}^{(i)}_R \star 
  (\rho_{\bar Q}(\widehat{\bar\Phi})^\dagger
  \star \widehat Q^{(j)}_L) \bigg )   \nonumber \\ 
&& - \sum_{i,j=1}^3  \bigg(
  G_d^{ij} ( \bar{ \widehat Q}^{(i)}_L \star \rho_Q(\widehat \Phi))\star  
  \widehat d^{(j)}_R
  + {G_d^\dagger}^{ij} \bar{ \widehat d}^{(i)}_R \star (\rho_Q(\widehat \Phi)^\dagger
  \star \widehat Q^{(j)}_L) \bigg ) \bigg)  \nonumber
\end{eqnarray}
We have to discuss three serious problems, namely
\begin{itemize}
\item[] the tensor product of gauge groups,
\item[] the so-called charge quantisation problem in nc QED,
\item[] the gauge invariance of the Yukawa couplings and 
\item[] ambiguities in the choice of the kinetic terms for the gauge fields in
the Lagrangian (\ref{b.1}).
\end{itemize}

\subsection{\label{sec5.1}Tensor Product of Gauge Groups}

There are several possibilities to deal with the tensor product of gauge groups
which correspond to the freedom in the choice of SW maps. The most symmetric and
natural choice is to take the classical tensor product 
\be\label{d.11}
V_\mu = g' A_\mu (x) Y + {g\over 2} \sum_{a=1}^3 B_{\mu a}\sigma^a +
	{g_S\over 2} \sum_{a=1}^8 G_{\mu a} \lambda^a,
\ee
defining one overall, "master" gauge field $V_\mu$, with gauge paramter
\be
{\Lambda} = g' \alpha (x)Y+{g\over 2} \sum_{a=1}^{3} \alpha^L_{a}(x) \sigma^a
	+{g_S\over 2} \sum_{b=1}^{8} \alpha^S_{b}(x) \lambda^b.
\ee
The nc gauge field $\widehat V[V]$ and gauge parameter 
$\widehat \Lambda[\Lambda,V]$ are given by 
\bea
\widehat V_\xi[V] & = & V_\xi 
	+ \frac{1}{4} \theta^{\mu\nu}\{V_\nu,\partial_\mu V_\xi\} + 
	\frac{1}{4} \theta^{\mu\nu}\{F_{\mu\xi},V_\nu\} \\
&& + \mathcal{O}(\theta^2),\\
\label{e.2}
\widehat\Lambda & = & \Lambda 
	+ \frac{1}{4} \theta^{\mu\nu}\{V_\nu,\partial_\mu \Lambda\}+ 
	\mathcal{O}(\theta^2).
\eea
As a consequence, the gauge groups mix in higher order of $\theta$ and cannot 
be viewed independently.

\noindent Let me also say a few words on the general tensor product of two gauge
groups \cite{Aschieri:2002mc}. The most general solution of the gauge
consistency condition (\ref{a.10}) - for one gauge group - is given by
\be
\widehat \Lambda [A] = \Lambda + {1\over 2} \theta^{\mu\nu} \{ A_\nu, \pat_\mu
\Lambda \}_c + \mathcal O(\theta^2),
\ee
where
\be
\{ A, B \}_c := {1\over 2} \{ A, B \} + (c-1/2) [A,B],
\ee
$c$ is a complex function of space-time. We have $\{A,B\}_{1/2}={1\over 2} \{A,B\}$.
The according gauge field is of the form
\be
\widehat A_\mu[A] = A_\mu + {1\over 2} \theta^{\nu\sigma} \{ A_\sigma, \pat_\nu
A_\mu \}_c + {1\over
2} \theta^{\nu\sigma} \{ F_{\nu\mu}, A_\sigma\}_c + \mathcal O(\theta^2).
\ee 
The gauge parameter $\widehat \Lambda_{(\Lambda,\Lambda')}[A,A']$ of the 
tensor product of two gauge groups $G$ and $G'$ consists of two parts
\be
\widehat \Lambda_{(\Lambda,\Lambda')}[A,A'] = \widehat \Lambda_\Lambda[A,A'] +
\widehat \Lambda'_{\Lambda'}[A,A'],
\ee
because of the linearity in the classical case. $\widehat
\Lambda_{(\Lambda,\Lambda')}[A,A']$ has to fulfill consistency relation
(\ref{a.10}). Therefore both, $\widehat \Lambda_\Lambda[A,A']$ and $\widehat
\Lambda'_{\Lambda'}[A,A']$ have to satisfy (\ref{a.10}) by their own, and there
is an additional cross relation
\be
[\widehat \Lambda_\Lambda\stackrel{*}{,} \widehat \Lambda'_{\Lambda'}] +
i\delta_\Lambda \widehat \Lambda'_{\Lambda'} - i\delta_{\Lambda'} \widehat
\Lambda_\Lambda = 0.
\ee
The solution is given by
\bea\label{e.1}
\widehat \Lambda_{(\Lambda,\Lambda')}[A,A'] & = & \Lambda + \Lambda' + {1\over
  2}\theta^{\mu\nu} \Big(
  \{ A_\nu, \pat_\mu\Lambda \}_c + \{ A'_\nu, \pat_\mu \Lambda' \}_d \Big)\\
&& + (1-{\gamma(x)\over 2})\, \theta^{\mu\nu} A'_\nu \pat_\mu \Lambda +
{\gamma(x)\over 2} \theta^{\mu\nu}
A_\nu \pat_\mu \Lambda' + \mathcal O(\theta^2).\nonumber
\eea
$\gamma(x)$ is a real function and $c-1/2$ and $d-1/2$ purely imaginary. Solving
eqns. (\ref{a.11}) and (\ref{a.12}) using (\ref{e.1}) will provide us with 
the SW map for matter fields and the gauge field. The symmetric choice in
(\ref{e.2}) is recovered by choosing $\gamma=1$ and $c=d=1/2$.

\subsection{\label{5.2}Charge Quantisation in NCQED}

It seems that in  ncQED only charges $\pm q,0$ can be accounted for, 
once $q$ is fixed \cite{Hayakawa}. 
\be
\widehat{\mathcal D}_\mu \widehat\psi = \partial_\mu\widehat \psi - 
	i q \widehat A_\mu*\widehat\Psi
\ee
the only other couplings of the field $\widehat A_\mu$ to a matter field
consistent with the nc gauge transformation $\widehat\delta_\alpha \widehat
A_\mu= \partial_\mu \widehat\alpha + i[\widehat\alpha\stackrel{*}{,}\widehat A_\mu]$ are
\bea
\label{d.12}
\widehat{\mathcal D}^-_\mu\widehat\psi^- & = & \partial_\mu \widehat\psi^- + 
	i q \widehat\psi^- * \widehat A_\mu,\\
\label{d.13}
\widehat{\mathcal D}^0_\mu\widehat\psi^0 & = & \partial_\mu\widehat\psi^0,\\
\label{d.14}
\widehat{\mathcal D}^{0'}_\mu\widehat\psi^{0'} & = & \partial_\mu \widehat\psi^{0'}
	-i q [\widehat A_\mu\stackrel{*}{,}\widehat\psi^{0'}].
\eea
Other charges $q^{(n)}$ cannot be absorbed into the respective field
$\widehat A_\mu$, because of the commutator in
\bea
\widehat F_{\mu\nu} & = & \partial_\mu \widehat A_\nu - 
	\partial_\nu \widehat A_\mu
	+ieq [\widehat A_\mu\stackrel{*}{,}\widehat A_\nu],\\
\widehat \delta_\Lambda \widehat A_\mu & = & \pat_\mu \widehat \Lambda + i [\widehat
	\Lambda \stackrel{*}{,} \widehat A_\mu].
\eea
Classically, we can have two particles $\psi$ and $\psi'$ with charges $q$ and
$q'$ coupling to the same gauge field. The gauge transformation of these fields
has the form
\beb
\delta\psi & = & iqe\lambda \psi, \\
\delta\psi' & = & iq' e\lambda'\psi',
\eeb
with covariant derivatives
\beb
\mathcal D_\mu\psi & = & \partial_\mu \psi - ieqa_\mu\psi,\\
\mathcal D_\mu\psi' & = & \partial_\mu \psi' - ieq'a'_\mu\psi',
\eeb 
where 
\beb
\delta a_\mu & = & \pat_\mu \lambda,\\
\delta a'_\mu & = & \pat_\mu \lambda'.
\eeb
Now, let us assume $\lambda'=\lambda$. We can consistently define
$$
a_\mu = {q'\over q} a'_\mu, \qquad f_{\mu\nu} = {q'\over q}f'_{\mu\nu}.
$$
The $*$-commutators spoil this simple picture.
The solution is again provided by SW maps. We have to introduce a different
gauge field $\widehat a^{(n)}_\mu$ for each distinct charge $q^{(n)}$ that appears 
in the theory. It seems that there are too many degrees of freedom, but the
SW map for $\widehat a^{(n)}_\mu$ is an expansion in the commutative gauge field
$a_\mu$ and $\theta$ only,
\be
\widehat a^{(n)}_\mu = a_\mu + {eq^{(n)}\over 4}\theta^{\sigma\tau}
	\{\partial_\sigma a_\mu , a_\tau\}
+ {eq^{(n)}\over 4} \theta^{\sigma\tau} \{f_{\sigma\mu}, a_\tau\} +
	\mathcal{O}(\theta^2).
\ee
The degrees of freedom are reduced to the classical ones.

\subsection{\label{sec5.3}Yukawa Couplings}

Let us now consider the Yukawa coupling terms in (\ref{b.1}) and their behaviour
under gauge transformation. They involve products of three fields, e.g.,
\be\label{e.3}
 -\sum_{i,j=1}^3  \bigg(
  W^{ij} ( \bar{ \widehat L}^{(i)}_L \star \rho_L(\widehat \Phi))
  \star  \widehat e^{(j)}_R
  + {W^\dagger}^{ij} \bar {\widehat e}^{(i)}_R \star (\rho_L(\widehat \Phi)^\dagger \star \widehat
  L^{(j)}_L) \bigg )
\ee 
Only in the case of commutative space-time, $\Phi$ commutes with generators of
$U(1)$ and $SU(3)$ groups. Therefore the Higgs field needs to transform from 
both sides, in
order to "cancel charges" from fields on either side (e.g.,
$\bar{ \widehat L}^{(i)}_L$ and $\widehat e^{(j)}_R$ in (\ref{e.3})).
The expansion of $\widehat\Phi$ transforming on the left and on the right under arbitrary
gauge groups is called hybrid SW map,
\bea
&\widehat\Phi[\Phi,A,A']  =  \phi + \frac{1}{2}\theta^{\mu\nu} A_\nu
  \Big(\partial_\mu\phi -\frac{i}{2} (A_\mu \phi + \phi A'_\mu)\Big)&\\
&  - \frac{1}{2}\theta^{\mu\nu} 
 \Big(\partial_\mu\phi - \frac{i}{2} (A_\mu \phi + \phi A'_\mu)\Big)A'_\nu 
 + \mathcal{O}(\theta^2),&
\eea
with $\widehat \delta \widehat\Phi = i\widehat\Lambda*\widehat\Phi-i\widehat\Phi*
\widehat\Lambda'$.
In the above Yukawa term (\ref{e.3}), we have $\rho_L(\widehat \Phi)= \widehat
\Phi[\phi,V,V']$, with
\beb
V_\mu & = & - {1\over 2}g' \mathcal A_\mu + g B_\mu^aT_L^a, \\ 
V'_\mu & = & g' \mathcal A_\mu.
\eeb
We further need a different representation for $\widehat\Phi$ in each of the
Yukawa couplings 
\bea
\rho_Q(\widehat\Phi) & = & \widehat \Phi [\phi,\,\frac{1}{6} g' {\cal A}_\mu + 
   g B^a_\mu T^a_L + g_S G_\mu^a T^a_S,\,
   \frac{1}{3} g' {\cal A}_\nu - 
   g_S G_\nu^a T^a_S  ],\\
\rho_{\bar Q}(\widehat\Phi) & = & \widehat\Phi[\phi,\,\frac{1}{6} g' 
  {\cal A}_\mu + g B^a_\mu T^a_L + g_S G_\mu^a T^a_S,\,-\frac{2}{3} g' {\cal A}_\nu - g_S
  G_\nu^a T^a_S].  
\eea
The respective sum of the gauge fields on both sides gives the proper quantum
numbers of the Higgs shown in Table \ref{table1}.

\subsection{\label{sec5.4}Kinetic Terms for the Gauge Bosons}

As we have mentioned earlier in Section \ref{sec3.1}, the kinetic terms for the
gauge field in the classical theory are determined uniquely by the requirements
of gauge invariance and renormalisability. In the non-commutative case, we
do not have a principle like renormalisability at hand. Gauge invariance alone
does not fix these terms in the Lagrangian. The non-commutative Standard Model as 
defined here, has rather to
be considered as an effective theory, where renormalisability is not applicable.
Otherwise, the role of the non-commutativity $\theta$ has to be considered very
carefully. $\theta$ may become a space-time field with a kinetic term of its own. 
Therefore, the representations used in the trace of the kinetic terms for the 
gauge bosons are not uniquely determined.
We will take the simpliest choice, since we are interested in a version of the
Standard Model on non-commutative space-time with minimal modifications. This
choice is discussed in Subsection \ref{sec5.4.1}. In Subsection \ref{sec5.4.2} we
will consider a maybe more physical and natural choice of representation.
Considering a Standard Model originating from a $SO(10)$ GUT theory 
\cite{Aschieri:2002mc}, these terms have a unique non-commutative generalisation.

\subsubsection{\label{sec5.4.1}Minimal Non-Commutative Standard Model}

The simplest choice for the gauge kinetic terms is named minimal Non-Commutative
Standard Model. The gauge kinetic terms
have the form displayed in eqn. (\ref{b.1}),
$$
-\int d^4x \frac{1}{2 g'} 
  \mbox{{\bf tr}}_{\bf 1} \widehat
  F_{\mu \nu}  *   \widehat F^{\mu \nu}
  -\int d^4x \frac{1}{2 g} \mbox{{\bf tr}}_{\bf 2} \widehat
  F_{\mu \nu}  *   \widehat F^{\mu \nu}
  -\int d^4x \frac{1}{2 g_S} \mbox{{\bf tr}}_{\bf 3} \widehat
  F_{\mu \nu}  *   \widehat F^{\mu \nu}.
$$
It is the sum of traces over the $U(1)_Y$, $SU(2)_L$ and $SU(3)_C$ sectors.
$\mbox{{\bf tr}}_{\bf 2}$ and $\mbox{{\bf tr}}_{\bf 3}$ are the usual
$SU(2)_L$ and $SU(3)_C$ traces, respectively. $\mbox{{\bf tr}}_{\bf 1}$ is the 
trace over the $U(1)_Y$ sector with 
$$
Y = {1\over 2} \left( \begin{array}{cc} 1 & 0 \\ 0 & -1
					   \end{array} \right)
$$
as representation of the charge generator.

\subsubsection{\label{sec5.4.2}Non-Minimal Non-Commutative Standard Model}

A perhaps more physical version of the Non-Commutative Standard Modell is obtained, 
if we consider a charge
matrix $Y$ containing all the fields of the Standard Model with covariant
derivatives acting on them. For the simplicity of presentation we will only
consider one family of fermions and quarks. Then the charge matrix has the form
\be\label{e.3a}
Y = \left( \begin{array}{cccccccc} -1 &&&&&&& \\ & -1/2 &&&&&& \\
	&& -1/2 &&&&& \\ &&& 2/3 &&&& \\ &&&& 2/3 &&& \\ &&&&& 2/3 & \\ 
	&&&&&& -1/3 & \\ &&&&&&& \ddots \end{array} \right),
\ee					   
according to Table \ref{table1}. It acts on fields given by column vectors
containing all the particles of the theory,
\be\label{e.4}
\Psi = \left( \begin{array}{c}e_R\\L_L\\u_R\\d_R\\Q_L\\\phi\end{array}\right).
\ee
The kinetic term for the gauge field is given by
\be
S_{\mbox{\tiny gauge}} = - \int d^4x\, \mbox{Tr}\, {1\over 2G^2} \widehat
F_{\mu\nu} * \widehat F^{\mu\nu},
\ee
where $\widehat F_{\mu\nu}=\pat_\mu \widehat V_\nu - \pat_\nu \widehat V_\mu -
i[\widehat V_\mu \stackrel{*}{,} \widehat V_\nu]$. The classical gauge field 
$V_\mu$ is the sum
$$
V_\mu = g' A_\mu (x) Y + {g\over 2} \sum_{a=1}^3 B_{\mu a}\sigma^a +
	{g_S\over 2} \sum_{a=1}^8 G_{\mu a} \lambda^a,
$$
where $Y$ is defined by (\ref{e.3a}), and $\widehat V_\mu=\widehat V_\mu[V]$. 
$G$ takes
the role of the coupling constant for the "master" gauge field $\widehat V_\mu$.
It is an operator commuting with all the generators of $(Y,T_L^a,T_c^b)$. It 
can be
expressed in terms of $Y$ and the six constants $g_1,\dots,g_2$ referring to 
the six multiplets (cf. Table \ref{table1}),
\beb
G e_R & = & g_1 e_R,\\
G L_L & = & g_2 L_L,\\
& \vdots &
\eeb
In the classical limit only three combinations of these six constants are 
relevant. They correspond to the usual coupling constants, $g'$, $g$, $g_S$,
\bea
&{\displaystyle \mbox{\bf Tr}{1\over \mathbf{G}^2}\, Y^2 = {1\over 2g'^2},}&\\
&{\displaystyle \mbox{\bf Tr}{1\over \mathbf{G}^2}\, T^a_L T^b_L = \delta^{ab} {1\over
	2 g^2},}&\\ 
&{\displaystyle \mbox{\bf Tr}{1\over \mathbf{G}^2}\, T^a_S T^b_S = \delta^{ab} {1\over
	2 g_S^2}.}& 
\eea
The above equations read, taking the charge matrix (\ref{e.3a}) into account
\bea
&{\displaystyle
{1 \over g_1^2} + {1\over 2 g_2^2} + {4\over 3 g_3^2} + {1\over 3 g_4^2} + {1
	\over 6 g_5^2} + {1\over 2 g_6^2} = {1\over 2g'^2},}&\\
&{\displaystyle
\frac{1}{g_2^2} + \frac{3}{g_5^2} + \frac{1}{g_6^2} = \frac{1}{g^2},}& \\
&{\displaystyle
\frac{1}{g_3^2} + \frac{1}{g_4^2} +\frac{2}{g_5^2} = \frac{1}{g_S^2}.}&
\eea
These three equations define for fixed $g'$, $g$, $g_S$ a three-dimensional simplex in the six-dimensional moduli space
spanned by  $1/g_1^2$, \ldots, $1/g_6^2$. The remaining three degrees of freedom become relevant at order $\theta$
in the expansion of the non-commutative action. Interesting are in particular the following traces
corresponding to triple gauge boson vertices:
\be
\mbox{\bf Tr} \frac{1}{\mbox{\bf G}^2} Y^3 
= -\frac{1}{g_1^2} - \frac{1}{4 g_2^2} + \frac{8}{9 g_3^2} - \frac{1}{9 g_4^2} +
\frac{1}{36 g_5^2} + \frac{1}{4 g_6^2},
\ee
\be
\mbox{\bf Tr} \frac{1}{\mbox{\bf G}^2} Y T_L^a T_L^b 
= \frac{1}{2}\delta^{ab}\left(-\frac{1}{2g_2^2} + \frac{1}{2g_5^2} +
\frac{1}{2g_6^2} \right),
\ee
\be
\mbox{\bf Tr} \frac{1}{\mbox{\bf G}^2} Y T_S^c T_S^d = \frac{1}{2}\delta^{cd}\left(
\frac{2}{3g_3^2} - \frac{1}{3g_4^2} +\frac{1}{3g_5^2} \right).
\ee
We may choose, e.g., to maximise  the traces over $Y^3$
and $Y T_L^a T_L^b$. This will give 
$1/g_1^2 = 1/(2 g'^2) - 4/(3g_S^2) -1/(2 g^2)$,  $1/g_3^2 = 1/g_S^2$, $1/g_6^2 = 1/g^2$,
$1/g_2^2 = 1/g_4^2 = 1/g_5^2 = 0$
and
\bea
& \mbox{\bf Tr} \frac{1}{\mbox{\bf G}^2} Y^3 =  -\frac{1}{2 g'^2} 
	+ \frac{3}{4 g^2} + \frac{20}{9 g_S^2},& \\
& \mbox{\bf Tr} \frac{1}{\mbox{\bf G}^2} Y T_L^a T_L^b = \frac{1}{4 g^2} 
	\delta^{ab}, & \\
& \mbox{\bf Tr} \frac{1}{\mbox{\bf G}^2} Y T_S^c T_S^d = \frac{2}{6 g_S^2}
	\delta^{cd}.&
\eea
In the scheme that we have presented in Section \ref{sec5.4.1} all three
traces are zero.
One consequence is that while non-commutativity does not {\it require}
a triple photon vertex, such a vertex is nevertheless consistent with  
non-commutativity.
It is important to note that the values of all three traces are bounded
for any choice of constants.

\subsection{\label{sec5.5}Higgs Mechanism}

In the leading order of the expansion in $\theta$, we obtain
\begin{eqnarray}
S_{\mbox{\tiny Higgs}} & = & \int d^4x\Bigg( (D^{SM}_\mu\phi)^\dagger D^{SM \mu}\phi
   -\mu^2 \phi^\dagger \phi
-\lambda (\phi^\dagger \phi) (\phi^\dagger \phi) \Bigg)
   \\
\nonumber &&
+ \int d^4x \Bigg ( (D^{SM}_\mu\phi)^\dagger
   \left( D^{SM \mu}\rho_0(\phi^1) + \frac{1}{2}
   \theta^{\alpha \beta} \partial_\alpha V^{\mu} \partial_\beta \phi 
 + \Gamma^\mu \phi \right)
\\ 
\nonumber && +
\left(D^{SM}_\mu \rho_0 (\phi^1) + \frac{1}{2}
   \theta^{\alpha \beta} \partial_\alpha V_\mu \partial_\beta \phi 
 + {\Gamma_\mu} \phi \right)^\dagger D^{SM \mu}\phi
\\
\nonumber &&
+\frac{1}{4} \mu^2
\theta^{\mu \nu} \phi^\dagger (g' f_{\mu \nu} + g F^L_{\mu \nu}) \phi
-  \lambda i \theta^{\alpha \beta}
\phi^\dagger \phi (D^{SM}_\alpha \phi)^\dagger (D^{SM}_\beta \phi)
\Bigg) + {\cal O}(\theta^2),
\end{eqnarray}
where
\be
\Gamma_\mu = - i V^1_\mu = \frac{i}{4}
    \theta^{\alpha \beta}\, 
    \{ g'{\cal A}_\alpha + g B_\alpha, \,\, g'\partial_\beta {\cal A}_\mu
    + g \partial_\beta B_\mu + g' f_{\beta \mu} + g F^L_{\beta \mu} \}.
\ee
We have also used the representation $\rho_0$, 
\begin{equation}
    \rho_0(\widehat \Phi)=\phi+\rho_0(\phi^1)+\mathcal{O}(\theta^2),
\end{equation}
where
\be
\rho_0(\phi^1) = - \frac{1}{2} \theta^{\alpha \beta}
      (g'{\cal A}_\alpha+g B_\alpha)\, \partial_\beta \phi
      + \frac{i}{8} \theta^{\alpha \beta}\,
      [ g'{\cal A}_\alpha+g B_\alpha, \, g'{\cal A}_\beta+g B_\beta] \, \phi.
\ee
For $\mu^2<0$ the $SU(2)_L\times U(1)_Y$ gauge symmetry is
spontaneously broken to $U(1)_{Q}$, which is the gauge group
describing the electromagnetic interaction. We have gauge invariance
and may choose the so-called unitarity gauge,
\begin{eqnarray}
\phi &=&\left(\begin{array}{c} 0 \\
     \eta +v \end{array} \right ){1\over \sqrt{2}},
\end{eqnarray}
where $v$ is the vacuum expectation value. Since the zeroth order of
the expansion of the non-commutative action corresponds to the
Standard Model action, the Higgs mechanism generates the same masses for
electroweak gauge bosons as in the commutative Standard Model,
\begin{eqnarray}
 M_{W^\pm}=\frac{g v}{2} \ \ \ \mbox{and} \ \ \
 M_Z=\frac{\sqrt{g^2+g'^2}}{2} v,
\end{eqnarray}
where the physical mass eigenstates $W^\pm$, $Z$ and $A$ are as
usual defined by
\begin{equation}
   W^\pm_\mu=\frac{B^1_\mu \mp i B^2_\mu}{\sqrt{2}}, \quad
   Z_\mu=\frac{-g'{\cal A}_\mu+gB^3_\mu}{\sqrt{g^2+g'^2}}
\quad \mbox{and} \quad
   {A}_\mu=\frac{g{\cal A}_\mu+g' B^3_\mu}{\sqrt{g^2+g'^2}}.
\end{equation}
There are no corrections to the masses of order $\theta$ since these terms
involve derivatives and therefore do not resemble mass terms.
The Higgs mass is given by $m^2_\eta=-2\mu^2$. Rewriting the
term $\Gamma_\mu$ in terms of the mass eigenstates, using
\begin{eqnarray}
B^3_\mu=\frac{g Z_\mu+g' A_\mu}{\sqrt{g^2+g'^2}} \ \ \ \mbox{and} \ 
\ \
   {\cal A}_\mu=\frac{g A_\mu-g' Z_\mu}{\sqrt{g^2+g'^2}},
\end{eqnarray}
one finds that besides the usual Standard Model couplings, numerous
new couplings between the Higgs boson and the electroweak gauge bosons
appear.

\subsection{\label{sec6}Conclusions and Remarks}

\begin{itemize}
\item The action of the NCSM (\ref{b.1}) has been expanded up to first order in 
the non-commutativity $\theta$ \cite{Calmet:2001na}. To $0^\textrm{th}$ order, 
we get the commutative standard model.

\item The different interactions cannot be considered seperately in
the NCSM, since we had to introduce the overall gauge field $V_\mu$ in 
(\ref{d.11}). They mix due to the SW maps, where the nc gauge field is given by
$\widehat V = \widehat V [V]$.

\item Higgs mechanism and Yukawa sector can be implemented in NCSM. The masses of
the gauge bosons and fermions equal those of the Standard Model on commutative 
space-time, at tree level.

\item There is no coupling of the Higgs to the electromagnetic Photon, 
in the minimal version. That coupling might be expected due to eqn. (\ref{d.14}).

\item Furtheron, no self interaction of the $U(1)_Y$ bosons occur, as one can see
from the expansion of the gauge kinetic terms. Vertices with five and six gauge 
bosons for $SU(3)_C$ and $SU(2)_L$ are present.

\item UV/IR mixing \cite{Minwalla:1999px} is not present due to $\theta$ 
expansion.

\item New decay modes for hadrons are found, $2$ Quarks-Gluon-Photon vertex 
occur.

\item One has to search for experimental signature for non-commutativity. 
One suggestion is to search for $Z\to\gamma\gamma$ decays in the non-minimal 
NCSM \cite{Behr:2002wx} or other decays forbidden by Lorentz symmetry 
\cite{Trampetic:2002eb}.

\item Other suggestions come from astrophysics. A weak limit on the 
non-commutative scale is proposed in \cite{Schupp:2002up}. The coupling of
neutrinos to photons implies new energy loss mechanisms in stars. This leads to
an estimate for the non-commutativity scale. In the context of 
$\kappa$-deformation, it has been argued to search for observable effects in 
time delays of high-energy $\gamma$ rays or neutrinos from active gallaxies, 
such as Markarian 142. These effects are due to modified dispersion relations, 
see e.g., \cite{kosmos}.

\end{itemize}

\end{document}